\documentclass{article}
\usepackage{epsfig}
\usepackage{amsmath,amssymb,amsfonts}
\usepackage{mathrsfs}
\usepackage{bbm}
\usepackage{cite}
\usepackage{slashed}
\usepackage{graphicx}
\usepackage{caption}
\usepackage{subcaption}
\usepackage{makecell}
\usepackage{multirow}
\usepackage{arydshln}
\usepackage{verbatim}
\usepackage[T1]{fontenc}
\usepackage[utf8]{inputenc}
\usepackage{ifthen}
\usepackage{multirow}
\usepackage{forloop}
\usepackage[english]{babel}
\usepackage{mathbbol}
\usepackage{xcolor}
\usepackage[a4paper, left=2cm, right=2cm, top=2cm, bottom=2cm]{geometry}

\usepackage{bbding}

\definecolor{darkgreen}{rgb}{0.2,0.6,0}
\definecolor{lightblue}{rgb}{0,0.5,0.8}
\definecolor{lightred}{rgb}{0.8,0.2,0.2}
\definecolor{darkorange}{rgb}{1,0.549,0}
\definecolor{pink2}{RGB}{255,128,128}
\definecolor{yellow2}{RGB}{255,255,0}
\definecolor{purple2}{RGB}{128,0,128}

\newcommand{\be}{\begin{equation}}
\newcommand{\ee}{\end{equation}}
\newcommand{\bea}{\begin{align}}
\newcommand{\eea}{\end{align}}

\newcommand{\p}{\partial}

\newcommand{\rph}{r_{\rm ph}}
\newcommand{\rsh}{r_{\rm sh}}

\usepackage[normalem]{ulem}

\allowdisplaybreaks

\title{Probing Quadratic Gravity with the Event Horizon Telescope}
\author{Jesse Daas$^1$, Kolja Kuijpers$^1$, Frank Saueressig$^1$, Michael F.\ Wondrak$^{2,3}$, and Heino Falcke$^2$\\[1ex]
\footnotesize
$^1$ Department of High Energy Physics/IMAPP, Radboud University, PO Box 9010,
6500 GL Nijmegen, The Netherlands\\
\footnotesize
$^2$ Department of Astrophysics/IMAPP, Radboud University, PO Box 9010,
6500 GL Nijmegen, The Netherlands\\
\footnotesize
$^3$ Department of Mathematics/IMAPP, Radboud University, PO Box 9010,
6500 GL Nijmegen, The Netherlands
}

\begin{document}

\maketitle

\begin{abstract}
Quadratic gravity constitutes a prototypical example of a perturbatively renormalizable quantum theory of the gravitational interactions. In this work, we construct the associated phase space of static, spherically symmetric, and asymptotically flat spacetimes. It is found that the Schwarzschild geometry is embedded in a rich solution space comprising horizonless, naked singularities and wormhole solutions. Characteristically, the deformed solutions follow the Schwarzschild solution up outside of the photon sphere while they differ substantially close to the center of gravity. We then carry out an analytic analysis of observable signatures accessible to the Event Horizon Telescope, comprising the size of the black hole shadow as well as the radiation emitted by infalling matter. On this basis, we argue that it is the brightness within the shadow region which constrains the phase space of solutions. Our work constitutes the first step towards bounding the phase space of black hole type solutions with a clear quantum gravity interpretation based on observational data.
\end{abstract}

\section{Probing Gravity with the Event Horizon Telescope}
General relativity constitutes the benchmark for describing gravitational phenomena. Its predictions have been confirmed in a vast number of experiments ranging from fundamental tests of the equivalence principle over the bending of light, the perihelion advance of mercury, and the Shapiro time delay, up to stellar system tests conducted within binary pulsar systems \cite{Will:2014kxa}. In the last decade, the direct detection of gravitational waves emitted during the coalescence of binary black hole systems \cite{abbott2016observation}, radio pulsar observations \cite{Wex:2020ald,2021}, and the image of a black hole shadow \cite{2000} taken by the Event Horizon Telescope (EHT) collaboration \cite{Akiyama:2019cqa} have opened a new era, by extending tests of general relativity into the strong gravity regime. Remarkably, the theory gives accurate predictions for these phenomena as well and there is currently no tension between theoretical forecasts and observations.  

From a theoretical perspective, there are compelling arguments that general relativity is not the ultimate theory describing gravity. A simple argument based on quantum field theory in a curved spacetime states that loop corrections involving ordinary matter fields induce new gravitational interactions beyond general relativity. At lowest order in perturbation theory these terms are four-derivative interactions built from squares of the spacetime curvature tensors. Typically, these corrections are suppressed by the Planck mass and therefore do not affect physics at macroscopic scales. Gravity can then be understood as an effective field theory where general relativity provides the leading terms capturing the low-energy physics.

This argument readily extends to the case where gravity itself is quantized. Starting from general relativity one finds that the theory is perturbatively non-renormalizable \cite{tHooft:1974toh,Goroff:1985th,vandeVen:1991gw,Bern:2015xsa}. At the same time the theory can be quantized consistently in the framework of an effective field theory \cite{Donoghue:1994dn}. Again, the leading correction terms encountered along this path are quadratic in the spacetime curvature \cite{Donoghue:2017pgk,Percacci:2017fkn}. Investigating the renormalizability of general relativity supplemented by terms quadratic in the curvature, it turns out that this quadratic gravity theory is perturbatively renormalizable \cite{Stelle:1976gc}, albeit at the prize of introducing additional ghost particles in its spectrum, see \cite{QGSalvio,Alvarez-Gaume:2015rwa} for recent discussions of this aspect. 

With the advent of observation channels probing gravity in the strong gravity regime, it is then natural to ask whether one can detect an imprint of gravitational higher-derivative interactions in the observational data. In this work, we investigate this question, analyzing potential signatures of quadratic gravity in the images made by the EHT. From the pioneering investigations \cite{Stelle:1977ry,Holdom:2002xy,Lu:2015psa,Lu:2015cqa, Lu:2015third} it is clear that quadratic gravity admits a rich solution space of geometries which behave like a Schwarzschild black hole sufficiently far away from the gravitational center but undergo severe modifications at scales where one would expect the event horizon of the black hole constructed within general relativity (see also \cite{Podolsky:2019gro} for a complete classification of possible scaling behaviors as well as exact solutions in quadratic gravity). These geometries comprise naked singularities as well as wormholes. A natural question to ask then is whether shadow observations can constrain the space of asymptotically flat vacuum solutions arising from the quadratic gravity theory.

In 2019, the EHT provided a unique test of general relativity by measuring the shadow size of the astrophysical black hole at the center of the galaxy M87$^\ast$ \cite{EventHorizonTelescope:2019ggy}. The Kerr solution relates this shadow size to the mass of the observed black hole which can also be determined by analyzing the motion of stars around it \cite{Gebhardt:2011yw} - thus allowing for a test of general relativity. A common misconception in popularized scientific literature is that the shadow is a direct consequence of the event horizon of the black hole. In fact, it is the impact parameter separating the light rays absorbed by the center of the geometry from the ones going out to infinity which determines its size, also see \cite{Bronzwaer:2021lzo} for a pedagogical account for the blocking and path-lengthening effects causing the shadow. It is worth mentioning that the region on the image plane that coincides with the shadow is also not completely dark owed to foreground emission: the shadow's brightness is just severely depressed compared to its surroundings, the latter exhibiting a  highly increased brightness due to the path-lengthening effect owed to the photon sphere. The location of the brightness depressed region turns out to be independent of astrophysical details of the matter surrounding and illuminating the black hole, as long as one remains in a realistic setting, again see \cite{Bronzwaer:2021lzo}. This is important, because otherwise it would not be possible to test general relativity. When looking at other observables measured by the EHT one therefore has to make sure that the emission model chosen does not influence the observables significantly, otherwise no statements can be made about the theory governing the underlying gravitational physics. With other gravitational theories in mind, it is then interesting to see what features are altered when creating the image using a different underlying gravitational theory, and whether or not these features correspond to observables that allow for gravitational tests. It has already been shown \cite{EventHorizonTelescope:2020qrl} that the parameters from the Parameterized Post-Newtonian formalism (artificially modifying the black hole metric) can be brought into contact with the EHT observations. Also constraints for concrete non-Kerr configurations and for general metric parametrizations can be derived, see \cite{Kocherlakota:2020kyu, Rezzolla:2014mua, Konoplya:2016jvv, Johannsen:2011dh, Johannsen:2013szh, Vigeland:2011ji}. Additionally in \cite{Eichhorn:2021etc, Eichhorn:2021iwq} observable imprints on the black hole image as a consequence of general requirements (e.g. regularity) have been investigated. For quantum gravity theories in particular however, new observational features for the EHT become irrelevant if modifications extend only to the Planck scale. This applies specifically to phenomenologically motivated regular black hole solutions \cite{Nicolini:2019irw,RBHBardeen,Hayward:2005gi, RBHFrolov, Dymnikova:1992ux,Bonanno:2000ep, Koch:2014cqa, Bonanno:2020fgp, Held:2019xde, Ashtekar:2005qt,Gambini:2013ooa, Knorr:2022kqp} where the image of the shadow is virtually indistinguishable from the one cast by the classical solutions once realistic parameters are chosen. In this light, possible signatures visible to the EHT must be associated with modifications of the geometry extending to macroscopic scales.

The rest of this work is organized as follows. We first discuss quadratic gravity and analyze its equations of motion in sect.\ \ref{sec.2}. In particular, sect.\ \ref{sec.22} discusses the long-distance behavior, and admissible scaling of solutions close to a specific value of the radius $r$. Sect.\ \ref{sec.23} then classifies the solutions by means of constructing the phase space. The observational imprints on EHT observations and the used emission model are discussed in sect.\ \ref{sec.3}. Finally, in the outlook we discuss several lines of research to follow up upon in the near-future. We work in units where $\hbar = c = G = 1$ throughout. We also neglect the non-analytic terms arising in the quantization of quadratic gravity at the one-loop level, leaving the inclusion of these genuine quantum effects for future work.

\section{Black Holes in Quadratic Gravity}
\label{sec.2}
In this section, we review the key properties of quadratic gravity in Sect.\ \ref{sec.21} before giving a local analysis of its vacuum solutions in Sect.\ \ref{sec.22}. The exposition of these sections mainly follows \cite{Lu:2015psa} where more details may be found. The novel result of this section is the phase space of global, static, spherically symmetric, and asymptotically flat vacuum solutions constructed in Sect.\ \ref{sec.23} (c.f.\ Fig.\ \ref{fig:pspace}). 
\subsection{Quadratic Gravity and spherically symmetric solutions}
\label{sec.21}
The action for quadratic gravity supplements the Einstein-Hilbert action by terms quadratic in the spacetime curvature
\be\label{eq:qg}
S = \frac{1}{16\pi}\int d^4x \sqrt{-g} \left[\gamma R - \alpha C_{\mu\nu\rho\sigma} C^{\mu\nu\rho\sigma} + \beta R^2 \right] \, . 
\ee
Here $R$ and $C_{\mu\nu\rho\sigma}$ are the Ricci scalar and the Weyl tensor constructed from the spacetime metric $g_{\mu\nu}$, $\gamma \equiv 1/G$, and $\alpha$, $\beta$ are two dimensionless coupling constants. The action \eqref{eq:qg} includes Einstein-Weyl gravity as a special case when setting $\beta=0$. Besides the massless graviton known from general relativity, the action \eqref{eq:qg} gives rise to a massive spin-two ghost and a massive spin-zero (non-ghost) excitation from the first and second quadratic term, respectively. In principle, one can also add a third term quadratic in curvature, the combination $E= R_{\mu\nu\rho\sigma} R^{\mu\nu\rho\sigma} - 4 R_{\mu\nu}R^{\mu\nu}+R^2$. In four spacetime dimensions $E$ constitutes the integrand of the Gauss-Bonnet term. Since this term is topological, it does not contribute to the equations of motion and will not be considered in the following. 

Varying \eqref{eq:qg} with respect to the spacetime metric gives the equations of motion
\be\label{eq:eom}
H_{\mu\nu} \equiv \gamma \left( R_{\mu\nu} - \frac{1}{2} g_{\mu\nu} R \right) 
- 4 \alpha \left(D^\rho D^\sigma + \frac{1}{2} R^{\rho\sigma} \right) C_{\mu\rho\nu\sigma}
+ 2 \beta \left( R_{\mu\nu} - \frac{1}{4} R g_{\mu\nu} - D_\mu D_\nu + g_{\mu\nu} D^2 \right) R = 0 \, , 
\ee
where $D_\mu$ is the covariant derivative and the symbol $H_{\mu\nu}$ is introduced for convenience.
Matter degrees of freedom can be included by adding the standard stress-energy tensor to the right-hand side of this equation. Since our work focuses on vacuum solutions, we do not consider this extension. Invariance of the action \eqref{eq:qg} under general coordinate transformations ensures the generalized conservation law
\be\label{eq:diffconst}
D^\mu \, H_{\mu\nu} = 0 \, . 
\ee
Furthermore, taking the trace of \eqref{eq:eom} gives
\be\label{eq:eomtrace}
\left( 6 \beta D^2 - \gamma \right) R = 0 \, . 
\ee
For $\beta = 0$ this relation implies $R=0$, stating that vacuum solutions of Einstein-Weyl gravity must have a vanishing Ricci scalar. In the presence of the $R^2$-coupling, this condition is not necessarily true and more general solutions are admissible.

We are interested in the static, spherically symmetric solutions of the system \eqref{eq:eom}. The most general spacetime metric compatible with these requirements can be cast into the form \cite{Hawking:1973uf}
\be\label{eq:metans}
ds^2 = - h(r) dt^2 + \frac{dr^2}{f(r)} + r^2 \left( d\theta^2 + \sin^2\theta d\phi \right) \, .
\ee
 The Schwarzschild metric is given by
\be\label{eq:ssmet}
h(r) = f(r) = 1 - \frac{2M}{r} \, ,
\ee
where $M$ is the ADM mass of the black hole according to Arnowitt, Deser, and Misner \cite{Arnowitt:1959ah}. 

In order to determine the functions $h(r)$ and $f(r)$, we substitute the ansatz \eqref{eq:metans} into the general equations of motion \eqref{eq:eom}. This leads to a set of coupled, non-linear differential equations. Out of those, only two are independent. We pick
\be\label{eq:dyn}
H_{rr} = 0 \, , \qquad H_{tt} = 0 \, . 
\ee
The angular equations are then fulfilled due to the constraint in eq.\ \eqref{eq:diffconst}. In the following, we focus on the generic case where $\alpha \not = 0$, $\alpha \not = 3\beta$, $\beta \not = 0$. In order to exhibit the differential order of the system \eqref{eq:dyn}, we follow \cite{Lu:2015psa}. We first observe that the highest order derivatives appearing in $H_{tt}$ are $f^{(3)}(r)$ and $h^{(4)}(r)$ while $H_{rr}$ depends on $f^{(2)}(r)$ and $h^{(3)}(r)$. We then define the auxiliary functions
\be
\begin{split}
X(r) = & \frac{h (\alpha -3 \beta ) \left(2 h r f' \left(2 h (\alpha +6 \beta )-r (\alpha -3 \beta ) h'\right)+f \left(12 h^2 (\alpha +6 \beta )-r^2
   (\alpha -3 \beta ) \left(h'\right)^2-4 h h' r (\alpha -3 \beta )\right)\right)}{\left(r (\alpha -3 \beta ) h'-2 h (\alpha +6 \beta
   )\right)^2}\\
Y(r) = & \,  \frac{2 r \, (\alpha - 3 \beta) f \, h^2}{2 (\alpha + 6 \beta) h - (\alpha - 3 \beta) r \, h^\prime} \, , 
\end{split} 
\ee
 and introduce the new equations
\be\label{eq:lincom}
\tilde{H}_{rr} \equiv H_{rr} \, , \qquad \tilde{H}_{tt} \equiv H_{tt} - X(r) H_{rr} - Y(r) \p_r H_{rr} \, . 
\ee
Clearly, the equations of motion \eqref{eq:dyn} are equivalent to
\be\label{eq:eom2}
\tilde{H}_{rr} = 0 \, , \qquad \tilde{H}_{tt} = 0 \, ,
\ee
i.e., we have rewritten the old equations of motion into an equivalent form which is tailored to exhibiting the order of the system. The first equation is of second order in $f$ and third order in $h$ while the second equation is of third order in $f$ and second order in $h$. Thus, taking the linear combination \eqref{eq:lincom} has reduced the differential order of the vacuum equation by one, which is remarkable since one generally expects fourth-order derivatives on $f$ and $h$ in a theory quadratic in curvature. In the present case, it is expected that, locally, solutions of the system \eqref{eq:eom2} are parameterized by six free parameters. 

\subsection{Local properties of the solutions}
\label{sec.22}
In order to construct the solutions of \eqref{eq:eom2}, we start by investigating the admissible asymptotic behaviors for large and small values of $r$.
\subsubsection{Asymptotic flatness}
\label{sec.221}
Let us start by analyzing the structure of the solutions for large values of $r$. In this regime, gravity is expected to be weak. Thus the asymptotics of the solutions can be obtained analytically by solving the linearized equations of motion obtained by perturbing the metric around flat Minkowski space. Following \cite{Bonanno:2021zoy}, this procedure is implemented by writing
\be
h(r) = 1 + \epsilon \, v(r) \, , \qquad f(r) = 1 + \epsilon \, w(r) \, . 
\ee
Substituting this expression into the non-linear equations \eqref{eq:eom2} and expanding to first order in $\epsilon$ then yields
\begin{align}\label{eq:linearized}
\begin{split}
& \left(4 \alpha -48 \beta -3 \gamma  r^2\right) w  \, -2 r^2 (\alpha -12 \beta ) w^{\prime\prime}
- r \left(4 \alpha + 24 \beta + 3 r^2 \gamma \right) v^\prime + 2 r^2 \left( \alpha +6 \beta  \right) \left( 2 v^{\prime\prime} + r v^{(3)} \right) = 0 \, , \\
&  r^3 \gamma \, (\alpha -3 \beta ) \left(2 v^\prime + r v^{\prime\prime} \right)
  -18 \alpha  \beta  \left( 2 w -2 r w^\prime +r^2 w^{\prime\prime} + r^3 w^{(3)}\right)  
  + r^2 \gamma (2 \alpha +3 \beta ) \left(w + r w^\prime \right) 
  = 0 \, . 
\end{split}
\end{align}
This system is readily solved using Fourier-methods. Introducing the masses of the massive spin-two and spin-zero degrees of freedom, 
\be\label{eq:masses}
(m_2)^2 = \frac{\gamma}{2 \alpha} \, , \qquad (m_0)^2 = \frac{\gamma}{6 \beta} \, , 
\ee
we get at linear order
\be\label{eq:linsol}
\begin{split}
h(r) = & \, 1 + C_T - \frac{2M}{r} + 2 S^+_2 \frac{e^{m_2 r}}{r} + 2 S_2^- \frac{e^{-m_2 r}}{r} + S^+_0  \frac{e^{m_0 r}}{r} + S^-_0 \frac{e^{-m_0 r}}{r} \, , \\
f(r) = & \, 1 - \frac{2M}{r} + S^+_2 \frac{e^{m_2 r}}{r} (1-m_2 r) +  S_2^- \frac{e^{-m_2 r}}{r} (1+m_2 r) - S^+_0  \frac{e^{m_0 r}}{r} (1-m_0 r) - S^-_0 \frac{e^{-m_0 r}}{r} (1+m_0 r) \, . 
\end{split}
\ee
Imposing that the  masses $m_0$ and $m_2$ are real restricts the values of the coupling constants to $\alpha \ge 0$, $\beta \ge 0$. This choice ensures the absence of oscillating factors in the linearized solution which are incompatible with asymptotic flatness. It is natural to expect that the masses are much smaller than those of astrophysical black holes.
The canonical normalization of the time-coordinate and asymptotic flatness fix the parameters $C_T =0$ and $S_2^+ = S_0^+ = 0$, respectively. Thus, for fixed $\alpha,\beta$, asymptotically flat solutions comprise a three-dimensional solution space spanned by $(M,S_0^-,S_2^-)$:
\be\label{eq:linsolflat}
\begin{split}
	h(r) = & \, 1 - \frac{2M}{r} + 2 S_2^- \frac{e^{-m_2 r}}{r} + S^-_0 \frac{e^{-m_0 r}}{r} \, , \\
	f(r) = & \, 1 - \frac{2M}{r} + S_2^- \frac{e^{-m_2 r}}{r} (1+m_2 r) - S^-_0 \frac{e^{-m_0 r}}{r} (1+m_0 r) \, . 
\end{split}
\ee
This expression admits a simple particle physics interpretation: for $S_2^- = S_0^- = 0$ the result for $h(r)$ and $f(r)$ agrees with the Schwarzschild solution. This part is governed by the massless degrees of freedom encoded in general relativity. The quadratic curvature terms lead to Yukawa-type corrections whose fall-off behavior at large distances is controlled by the spin-two mass $m_2$ and spin-zero mass $m_0$, respectively. The corrections to the Schwarzschild metric thus decrease exponentially as $r \rightarrow \infty$. Therefore, they escape the classification with the standard Frobenius method \cite{Saueressig:2021wam}.

For $\alpha = 0, \beta \not = 0$ and $\beta = 0, \alpha \not = 0$, the action integrand \eqref{eq:qg} reduces to the form $R + R^2$ and $R + C_{\mu\nu\rho\sigma} C^{\mu\nu\rho\sigma}$, respectively. From \eqref{eq:masses}, we observe that in these limits the masses diverge, $m_2 \rightarrow \infty$ and $m_0 \rightarrow \infty$, respectively, and the corresponding massive degrees of freedom decouple. At the level of \eqref{eq:linsolflat} this entails that the space of asymptotically flat solutions is spanned by only two free parameters, as either $S_2^-$ or $S_0^-$ drop out of the linearized solution. This can also be inferred from \eqref{eq:linearized}, where setting either $\alpha = 0$ or $\beta =0$ reduces the order of the linearized equations. We also note that the equal mass limit $m_0 = m_2$ (corresponding to the special case $\beta = \alpha/3$) does not reduce the dimension of the solution space, since the free parameters $S_2^-$ and $S_0^-$ enter into $h(r)$ and $f(r)$ in the form of two different linear combinations. Based on these insights, we will analyze the space of vacuum solutions for the generic situation where $\alpha \not =0$ and $\beta \not = 0$.

\subsubsection{Admissible scaling behavior for small $r$}
\label{sec:Analytic_small_r}
Upon constructing the solutions of the linearized system of equations, we now determine the admissible scaling behaviors as $r \rightarrow 0$. Later on, this will serve as a cross-check on the validity of our numerical solutions. We then write $h(r)$ and $f(r)$ as a power series in $r$, referred to as a Frobenius ansatz\footnote{Frobenius expansions are in general applicable to linear differential equations. In our investigations we follow the literature which extends the usage to non-linear differential equations, see for instance~\cite{Holdom:2002xy,Lu:2015psa}.}:
\be\label{eq:smallr}
h(r) = r^t \sum_{n=0}^\infty h_n \, r^n \, , \qquad f(r) = r^s \sum_{n=0}^\infty f_n \, r^n \, . 
\ee 
Here $h_0 \not = 0$ and $f_0 \not = 0$ by definition and the exponents $(s,t)$ capture the short-scale behavior of the solution. 
We then substitute the expansion \eqref{eq:smallr} into the system of non-linear equations \eqref{eq:eom2} and determine the admissible values for $s$ and $t$ by analyzing the equations appearing at leading order at $r=0$. The consistent values are referred to as analytic classes $(s,t)_0$ where the subscript ``0'' indicates that the expansion point is $r=0$. In this way, we confirm the three solutions given in \cite{Stelle:1976gc,Lu:2015psa}, which we tabulate in the upper half of Table \ref{tab.scaling1}. Going to higher powers of $r$ and solving the corresponding equations recursively for the highest coefficients $h_n, f_n$ contained in them allows to fix the power series in terms of a set of undetermined coefficients. For the admissible scaling behaviors, the number and choice of free coefficients is then listed in columns 4 and 5 of Table \ref{tab.scaling1}. Moreover, the scaling behaviors allow determining the strength of the curvature singularity by evaluating the Kretschmann scalar $R_{\mu\nu\rho\sigma}R^{\mu\nu\rho\sigma}$ in the limit $r \rightarrow 0$. The results are listed in the sixth column of Table \ref{tab.scaling1}, indicating that the naked singularity solutions are more divergent than the Schwarzschild solution.

\begin{table}[t!]
	\centering
\begin{tabular}{|c|c|c|c|l|c|}
	\hline
\makecell{Analytic class \\ $(s,t)_r$} & Literature \cite{Stelle:1976gc,Lu:2015psa} & Comments & \makecell{Number of \\ free parameters} & Free parameters & $R_{\mu\nu\rho\sigma}R^{\mu\nu\rho\sigma}$\\ \hline \hline
$ \bigg. (-2,2)_0$ & Stelle-$(2,2)$ & \makecell{naked singularity} & $6$ & $h_0$, $f_0$, $f_1$, $f_2$, $h_3$, $f_3$ & $\propto r^{-8}$  \\
\hdashline
$\bigg. (-1,-1)_0$ & \makecell{Stelle-$(1,-1)$ } & \makecell{Schwarzschild-like \\ naked singularity} & $4$ & $h_0, f_0, h_3, f_3$ & $\propto r^{-6}$ \\
 \hline
$\bigg. (0,0)_0$ & \makecell{Stelle-$(0,0)$} & \makecell{$f_0 = 1$ \\ non-singular} & $3$ & $h_0, h_2, f_2$ & finite  \\ 
\hline \hline
$\bigg. (0,0)_{r_0}$ & L\"u-$(0,0)_{r_0}$ & $r_0$: generic & $6$ & $\tilde{h}_0$, $\tilde{f}_0$, $\tilde{h}_1$, $\tilde{f}_1$, $\tilde{h}_2$, $\tilde{f}_2$ & finite \\
\hdashline
$\bigg. (1,1)_{r_0}$ &\makecell{L\"u-$(1,1)_{r_0}$} & $r_0$: horizon & $4$ & $r_0$, $\tilde{h}_0$, $\tilde{f}_0$, $\tilde{h}_1$ & finite \\
\hdashline
$\bigg. (1,0)_{r_0}$ &\makecell{L\"u-$(1,0)_{r_0}$} &  $r_0$: wormhole throat &  $3$ & $r_0$, $\tilde{h}_0$, $\tilde{f}_0$ & finite \\
\hline
\end{tabular}
\caption{\label{tab.scaling1} Summary of the admissible asymptotic scaling behaviors resulting from eq.\ \eqref{eq:smallr} (top block) and eq.\ \eqref{eq:Frobenius-r0} (bottom block). The latter entries are limited to the subset of the solutions \eqref{eq:scaling-families} relevant for our classification program. The results agree with earlier findings \cite{Stelle:1977ry,Holdom:2002xy,Lu:2015psa}.}
\end{table}

At this stage two technical remarks are in order. Firstly, the ansatz \eqref{eq:smallr} covers solutions which have a series expansion at $r=0$ only. Solutions which are not analytic at the expansion point would be missed by the scaling classification. Our numerical investigation did not provide evidence for the presence of such solutions, however, all of our solutions comply with the scaling behaviors tabulated in Table \ref{tab.scaling1}. Secondly, one still has the freedom of rescaling the time-coordinate by an arbitrary constant coefficient. This freedom could, in principle, be used to eliminate $h_0$ from the list of free parameters. When constructing global solutions, we have used this freedom to impose asymptotic flatness though. Hence this parameter is kept as a free parameter.

\subsubsection{Admissible scaling behavior around a finite radius $r_0$}
\label{sec. 223}
There are also spacetime solutions in which the asymptotic form of the metric at infinity does not allow a smooth continuation to $r=0$. These cases can be investigated by a Frobenius expansion at a finite radial coordinate $r_0$. 
Similarly to the ansatz~\eqref{eq:smallr}, one performs a formal power-series ansatz with integer values $n$
\be
\label{eq:Frobenius-r0}
h(r) 
= (r-r_0)^{t} \, \sum_{n=0}^\infty \tilde{h}_n (r-r_0)^{n} \, , \qquad 
f(r) 
= (r-r_0)^{s} \, \sum_{n=0}^\infty \tilde{f}_n (r-r_0)^{n} \, .
\ee
The possibility of having more general scaling laws including half-integer values for $n$ have been discussed in \cite{Lu:2015psa} but will not play a role in the following. Substituting the ansatz into the non-linear equations of motion, the admissible values $(s,t)_{r_0}$ and the series coefficients $\tilde{h}_n$ and $\tilde{f}_n$ (apart from the free parameters spanning the solution space) can be obtained by solving the resulting hierarchy of equations iteratively. The incidental polynomials, appearing at lowest order in this expansion, admit infinite families of solutions
\be\label{eq:scaling-families}
\begin{array}{ll}
	t = 0 \, , \qquad & s \le 1 \in \mathbb{Z} \, , \\
	t \ge 1 \in \mathbb{N} \, , \qquad & s = 2 - t \, .
\end{array}
\ee
A specific selection from these solutions, classified by the labels $(s,t)_{r_0}$, is given in the lower half of Table \ref{tab.scaling1}. The expansion at a regular point, where both $h(r_0)$ and $f(r_0)$ are finite, are covered by $(s,t)_{r_0} = (0,0)_{r_0}$. The horizon appearing in the Schwarzschild solution is encoded in the $(s,t)_{r_0} = (1,1)_{r_0}$ class. Moreover,  we encounter the class $(1,0)_{r_0}$, in which $f(r)$ vanishes linearly at $r_0$ while $h(r)$ stays finite. These solutions can be associated with wormhole solutions~\cite{Lu:2015psa}.

\subsection{Constructing the Phase Space}
\label{sec.23}
Eq.\ \eqref{eq:linsolflat} gives the parametrization of asymptotically flat solutions in the linearized regime. Fixing $\gamma = 1/G = 1$, which states that all dimensionful quantities are measured in Planck units, the solution space is spanned by five parameters: $\alpha$, $\beta$, $M$, $S_0^-$, and $S_2^-$. Evaluating the expression at $r_i > 3M$ then gives initial conditions for the non-linear equations of motion from which the solution can be integrated inward numerically.\footnote{Note that the process of extending the solutions to $r > r_i$ is computationally difficult for the following reason. At the linearized level, asymptotically flat solutions span a subspace of the full solution space \eqref{eq:linsol}. The additional modes excluded by the asymptotic flatness condition grow exponentially as $r$ increases. This makes the integration procedure unstable: tiny numerical errors can turn on the unstable perturbations which then grow and drive the solution away from asymptotic flatness. It is clear that this is a technical and not a conceptual problem though. Eq.\ \eqref{eq:linsolflat} guarantees that the asymptotically flat solutions extend to $r \rightarrow \infty$. Placing initial conditions further away from the center, it is then just a question of computational power to extend these conditions inward. Practically, we then chose initial conditions at $r_i = 3.5 M$ which we found a good working compromise between being in a regime where the Schwarzschild solution is a good approximation and the exponential corrections are still at a value which can be handled numerically.}

\begin{figure}[t!]
	\centering
	\begin{subfigure}{0.3\textwidth}
		\centering
		\includegraphics[width=\textwidth]{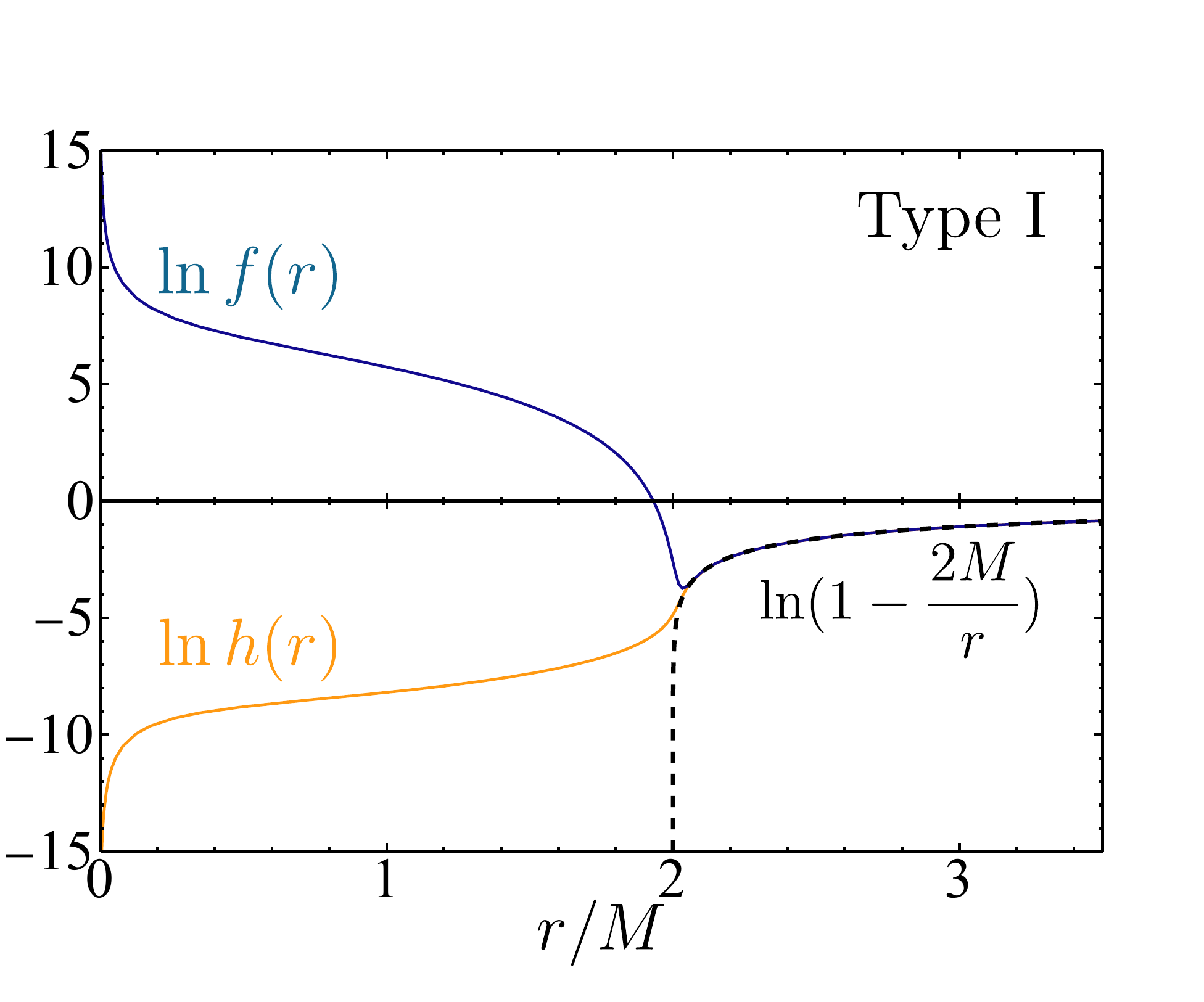}\\
		\includegraphics[width=\textwidth]{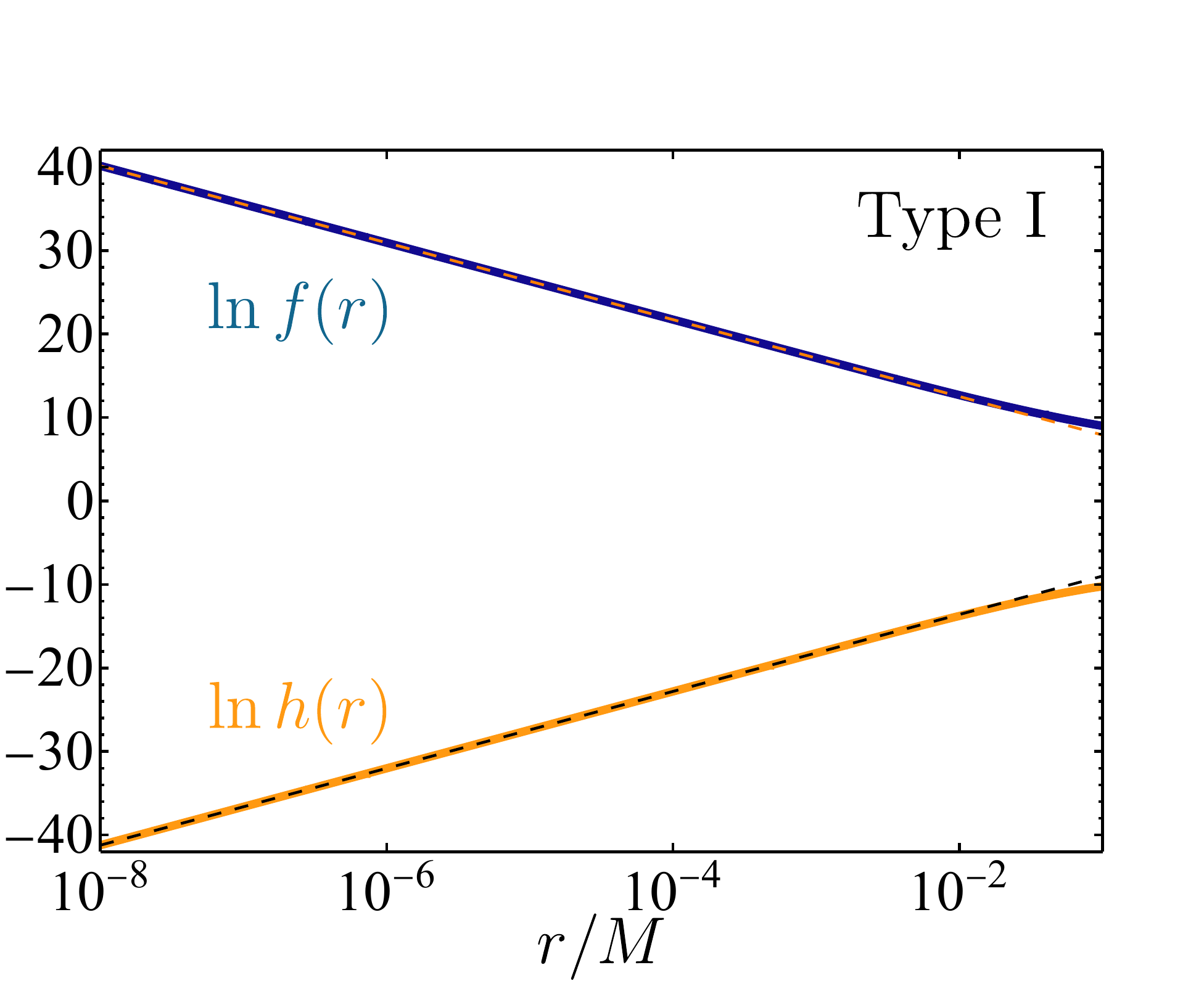}
		\caption{\label{fig:extype1}Type I}
	\end{subfigure}
	\hfill
	\begin{subfigure}{0.3\textwidth}
		\centering
		\includegraphics[width=\textwidth]{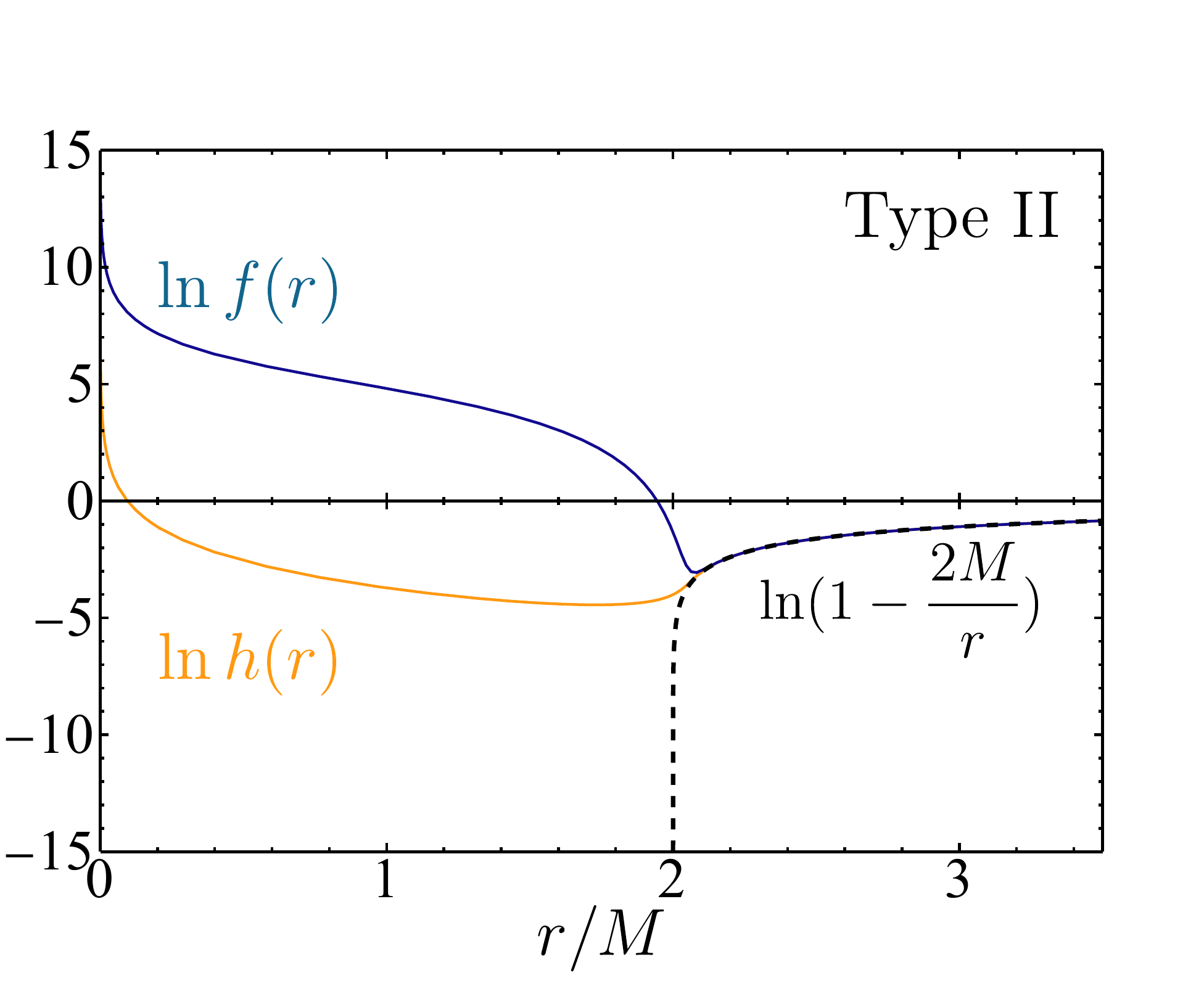}\\
		\includegraphics[width=\textwidth]{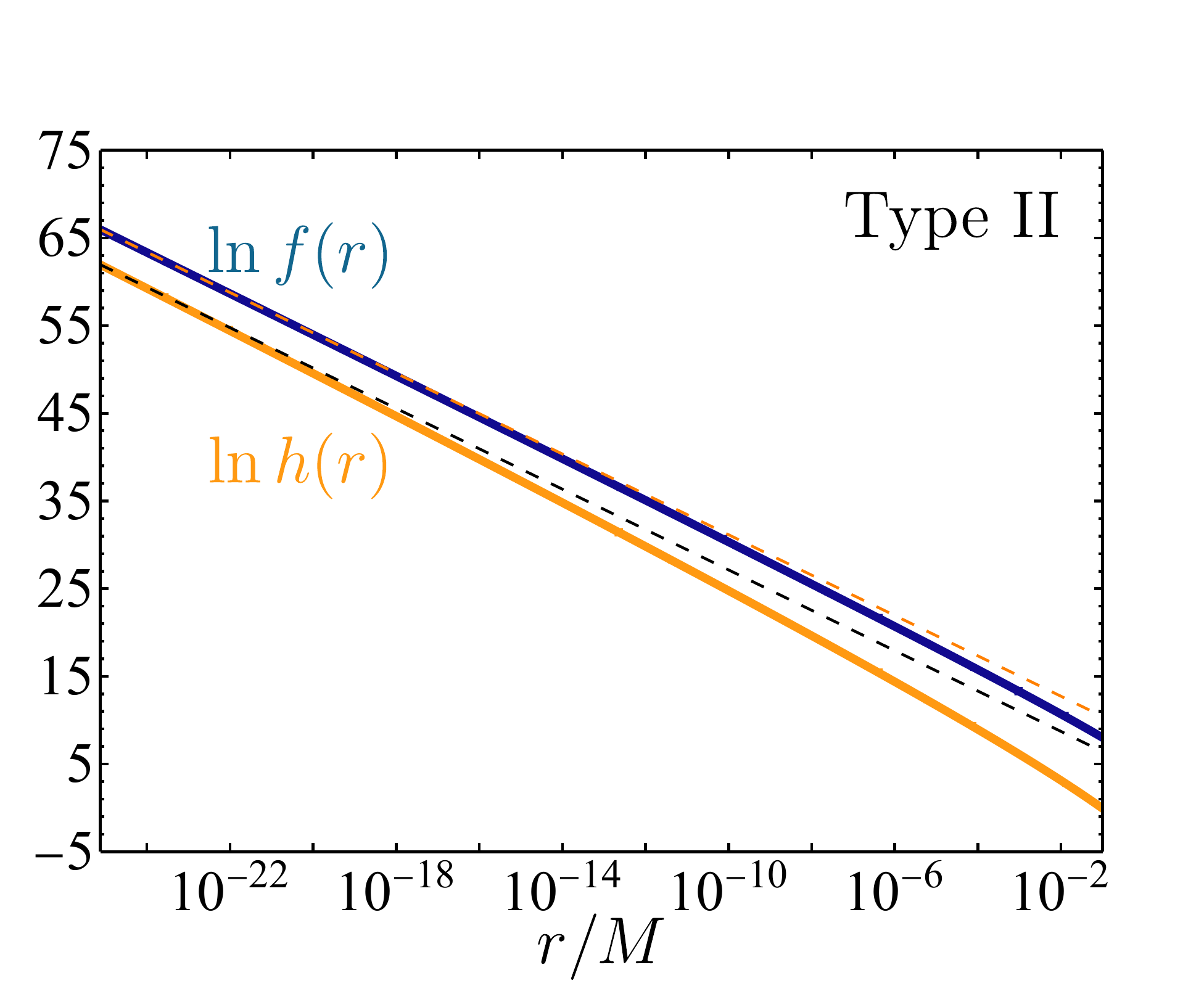}
		\caption{\label{fig:extype2}Type II}
	\end{subfigure}
	\hfill
	\begin{subfigure}{0.3\textwidth}
		\centering
		\includegraphics[width=\textwidth]{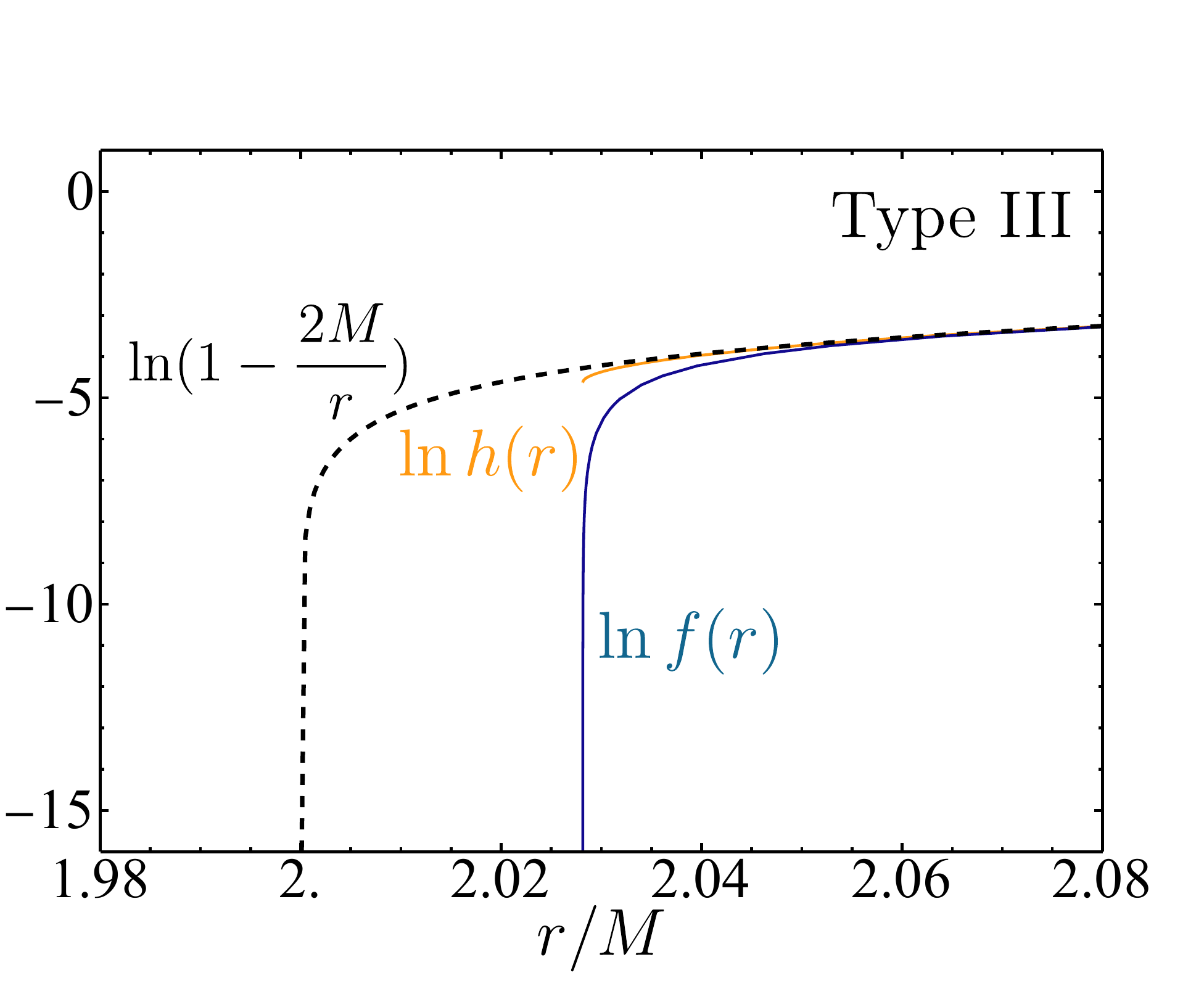}\\
		\includegraphics[width=\textwidth]{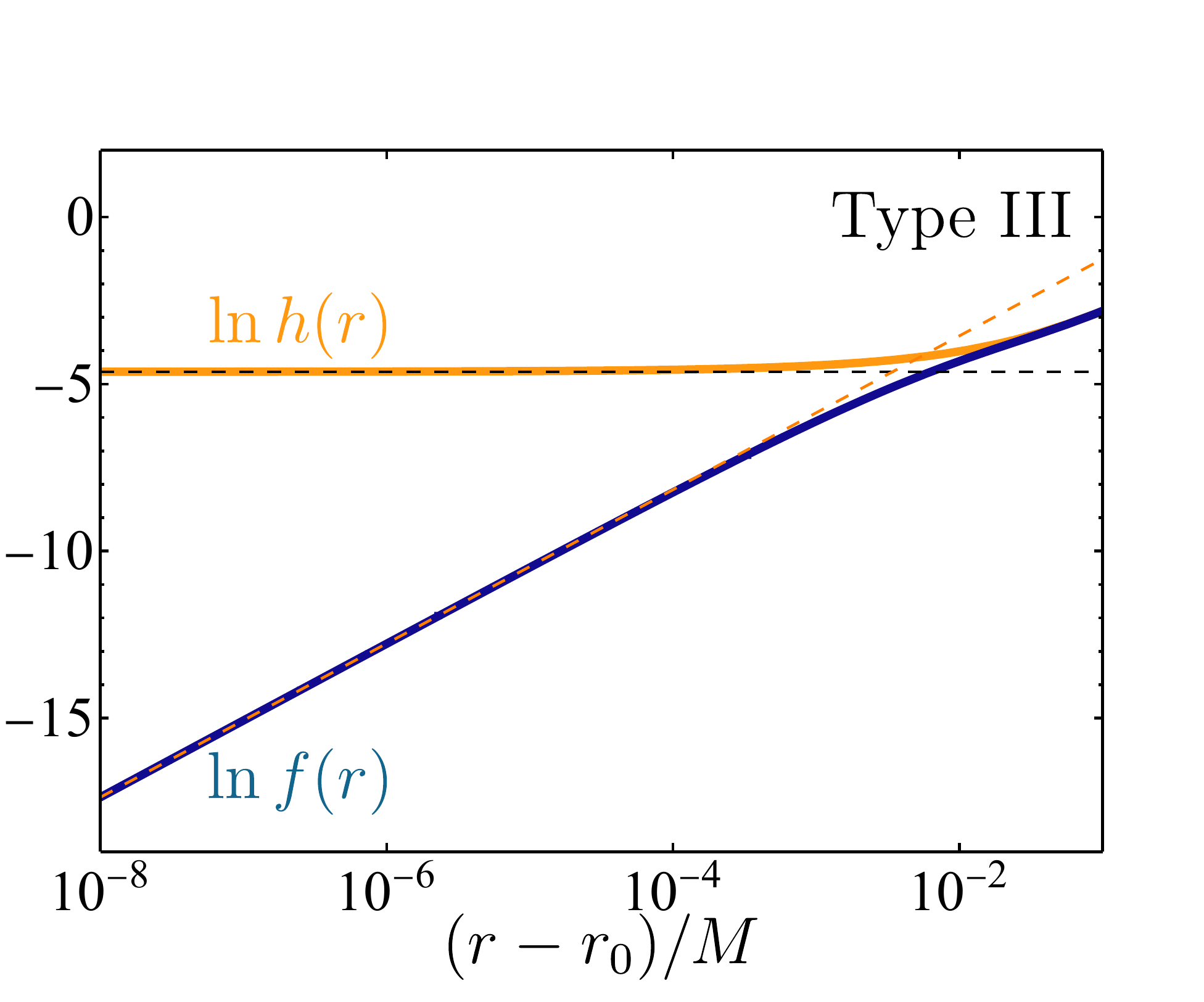}
		\caption{\label{fig:extype3}Type III}
	\end{subfigure}
	\caption{\label{fig.classexamples} Illustration of the topologically distinct global solutions in quadratic gravity. The solutions are obtained from solving \eqref{eq:eom2} numerically with initial conditions given in Table \ref{tab:initialconditions}. The top line displays the radial dependence of $\ln f$ (blue curve) and $\ln h$ (orange curve) for the solutions of (a) Type I, (b) Type II, and (c) Type III. The Schwarzschild solution with $M=10$ and an event horizon at $r/M=2$, is superimposed as the dashed line. The lower row shows the scaling behavior of the solutions as $r \rightarrow 0$ (Type I and Type II) as well as for $r - r_0 \rightarrow 0$ (Type III) in a double logarithmic presentation. The scaling behaviors are $(s,t)_0 = (-2,2)_0$ for Type I, $(s,t)_0 = (-1,-1)_0$ for Type II, and $(s,t)_{r_0} = (1,0)_{r_0}$ for Type III. Graphs of the respective power law are superimposed as dashed lines (black for $h$ and orange for $f$). This links the asymptotically flat, global solutions to the classification of local scaling behaviors identified in Table \ref{tab.scaling1}. }
\end{figure}

\begin{figure}
	\centering
	\begin{subfigure}{0.48\textwidth}
		\includegraphics[width=\textwidth]{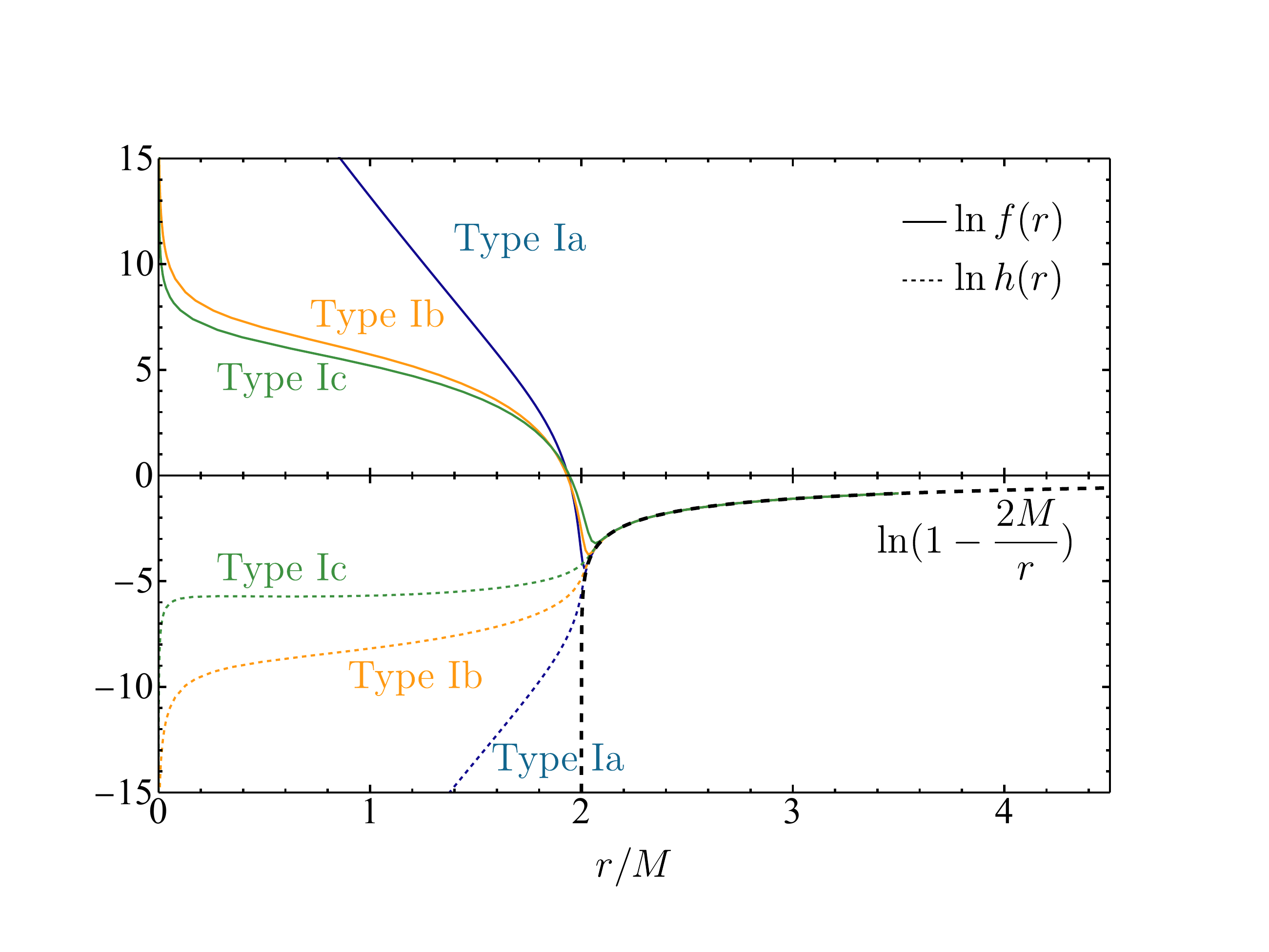}
	\end{subfigure}
	\hfill
	\begin{subfigure}{0.48\textwidth}
		\includegraphics[width=\textwidth]{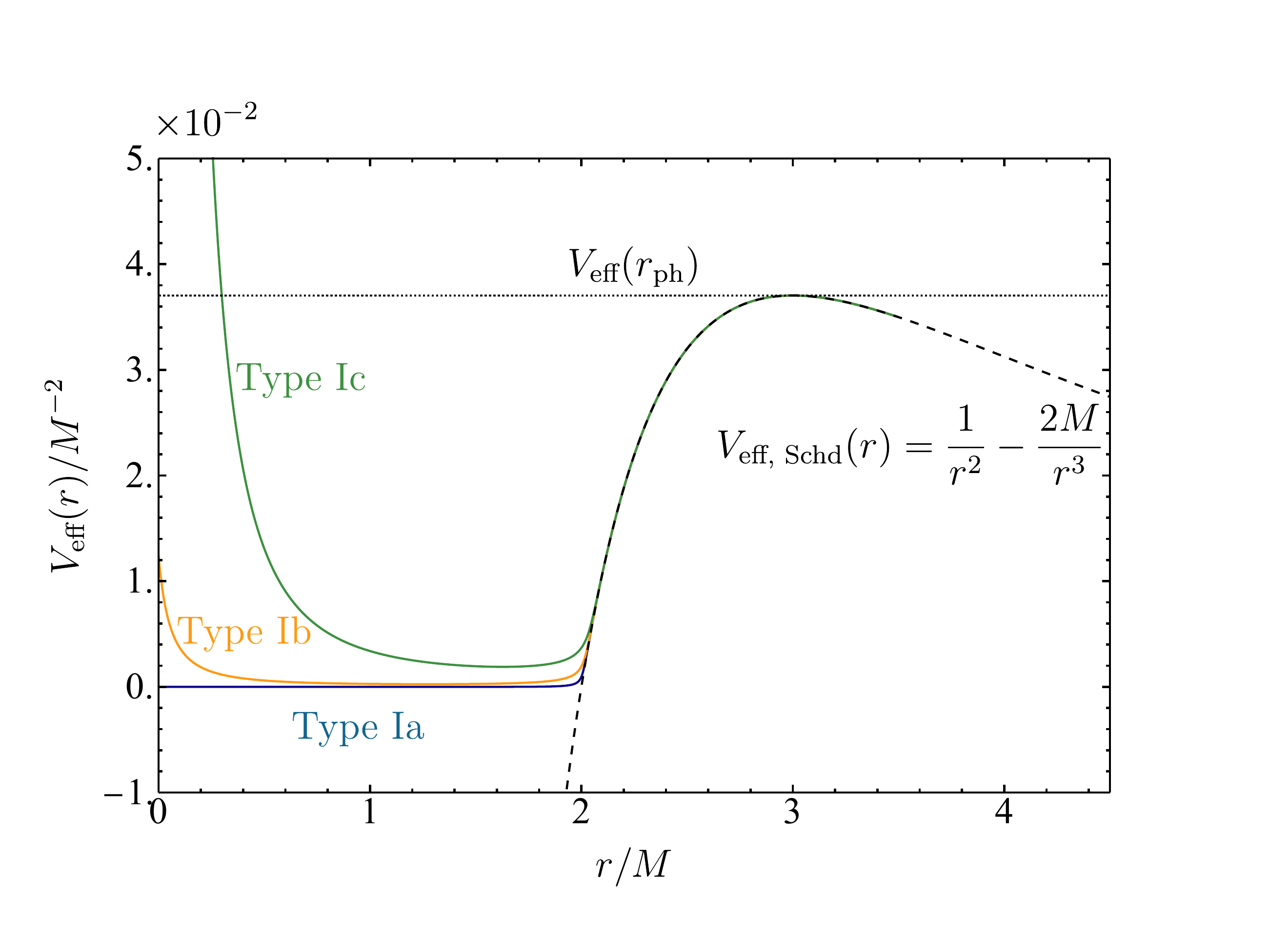}
	\end{subfigure} 
\caption{\label{fig:typeIref} Refinement of the Type-I classification. The solutions arise from the initial conditions given in Table~\ref{tab:initialconditions} for objects of mass $M=10$. The left panel displays $\ln f$ (top, solid) and $\ln h$ (bottom, dashed), showing the typical behavior of the Type I solutions given in Fig.\ \ref{fig:extype1}. The refinement follows from considering $V_{\rm eff} = h(r)/r^2$, shown in the right panel. For Type Ia solutions $V_{\rm eff}$ decreases monotonically for $r < 3M$. For Type Ib solutions the potential has a stable minimum and $V_{\rm eff}(0) < V_{\rm eff}(3M)$. The characteristic feature for Type Ic solutions is that $\infty > V_{\rm eff}(0) > V_{\rm eff}(3M)$.}
\end{figure}

By varying the parameters systematically, we have identified a number of topologically distinct solutions which can be discriminated according to their scaling behavior close to $r=0$ (or their termination point $r_{\rm term} > 0$, if applicable). The characteristic features of the resulting classes (Type I, II, and III) are then described below. This discussion is complemented by the representative examples shown in Figs.\ \ref{fig.classexamples} and \ref{fig:typeIref}, arising from the initial conditions listed in 
 Table \ref{tab:initialconditions}. Following \cite{Hernandez-Lorenzo:2020aie}, the scaling exponents are found by fitting to the global solutions in the vicinity close to the end-point. This procedure introduces a numerical uncertainty. In the case of Type I is almost perfect with a deviation of less than order $10^{-3}$. For Type II and Type III solutions we commonly observe deviations of approximately $5$\%.  For some Type III examples this may even increase to $50$\% in some extreme cases.
\begin{table}[t!]
	\centering
	\begin{tabular}{|c||ccc:ccc:c|cc|}
	\hline
	Solution class & $\gamma$ & $\alpha$ & $\beta$ & $M$ & $S_2^-$ & $S_0^-$ & $r_i$ & Example Fig & Intensity Fig\\
	\hline \hline
	Type I & 1 & 1/2 & 1/6 & 10 & 1/200 & -1/15 & 35 & \ref{fig:extype1} & \\
	\hdashline
	Type Ia & 1 & 1/2 & 1/6 & 10 & 1/200 & -1/150 & 35 & \ref{fig:typeIref} blue & \ref{fig:intss} \\
	Type Ib & 1 & 1/2 & 1/6 & 10 & 1/200 & -1/15 & 35 & \ref{fig:typeIref} orange & \ref{fig:intss} \\
	Type Ic & 1 & 1/2 & 1/6 & 10 & 1/200 & -2/3 & 35 & \ref{fig:typeIref} green & \ref{fig:inttype1c} \\
	\hdashline	
	Type II & 1 & 1/2 & 1/6 & 10 & 2 & 1/100 & 35 & \ref{fig:extype2} & \ref{fig:inttype2} \\
	Type III & 1 & 1/2 & 1/6	& 10 & 1/200 & 1/15 & 35 & \ref{fig:extype3} & \ref{fig:intss} \\
	\hline
	\end{tabular}
	\caption{\label{tab:initialconditions} Initial conditions for the numerical integration procedure generating the representative geometry for each class. All figures presented in our work build on the corresponding solutions.}
\end{table}
\paragraph{Type I:}
For these solutions the numerical integration is reliable until $r \approx 0$. For $r \gtrsim 2M$, the functions $h(r)$ and $f(r)$ essentially follow the Schwarzschild solution, before starting to deviate substantially for $r \lesssim 2M$. At $r \ll 1$, the scaling behavior of $h(r)$ and $f(r)$ follows the local expansion with $(s,t)_0 = (-2,2)_0$, i.e.,
\be\label{eq:hpowerlaw}
 f(r) \sim f_0 \, r^{-2} \, , \qquad h(r) \sim h_0 \, r^2  \, ,
\ee
where $\sim$ indicates that the relations hold asymptotically.  This constitutes the defining criterion for geometries of Type I. The characteristic example for these solutions is shown in Fig.\ \ref{fig:extype1}.

\begin{figure}[t!]
	\centering
	\begin{tabular}{|c|ccc|}
		\hline
		& \multicolumn{3}{|c|}{$S_{0}^{-}$} \\
		\hline
		& \makecell{\rule{0pt}{29ex}\includegraphics[scale=0.60]{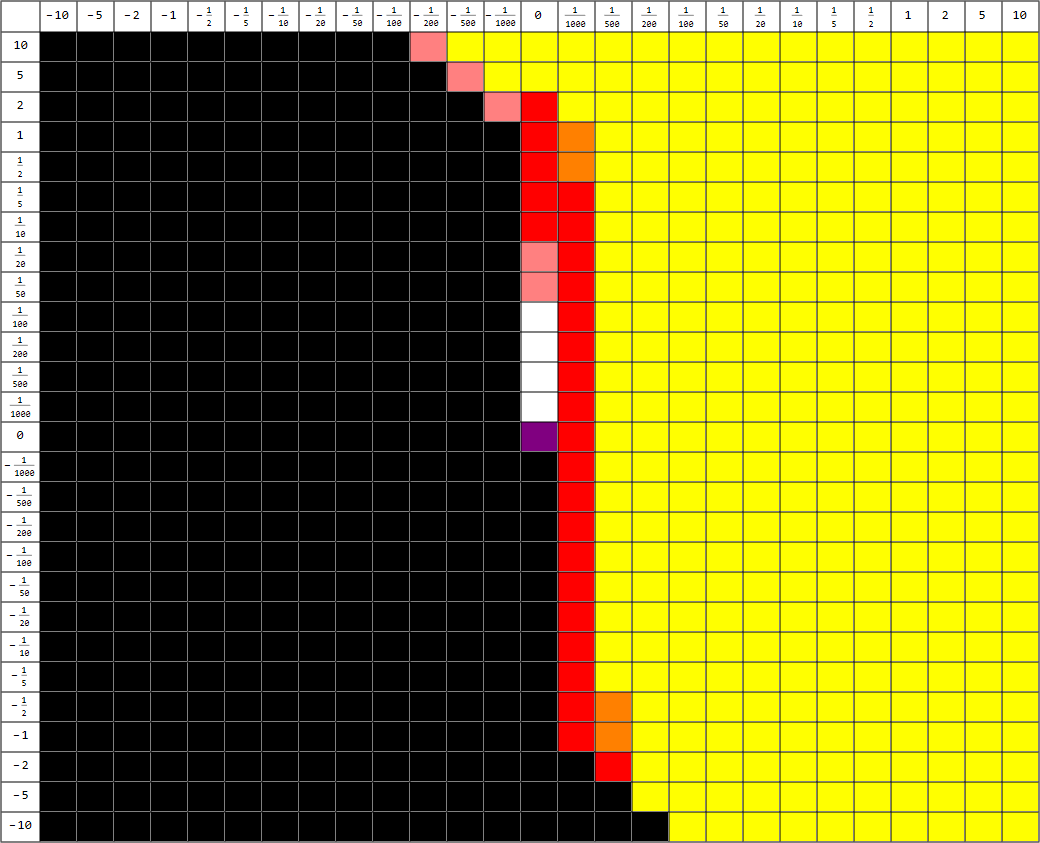} \\ $\beta=1/3$ } & \makecell{\rule{0pt}{29ex}\includegraphics[scale=0.60]{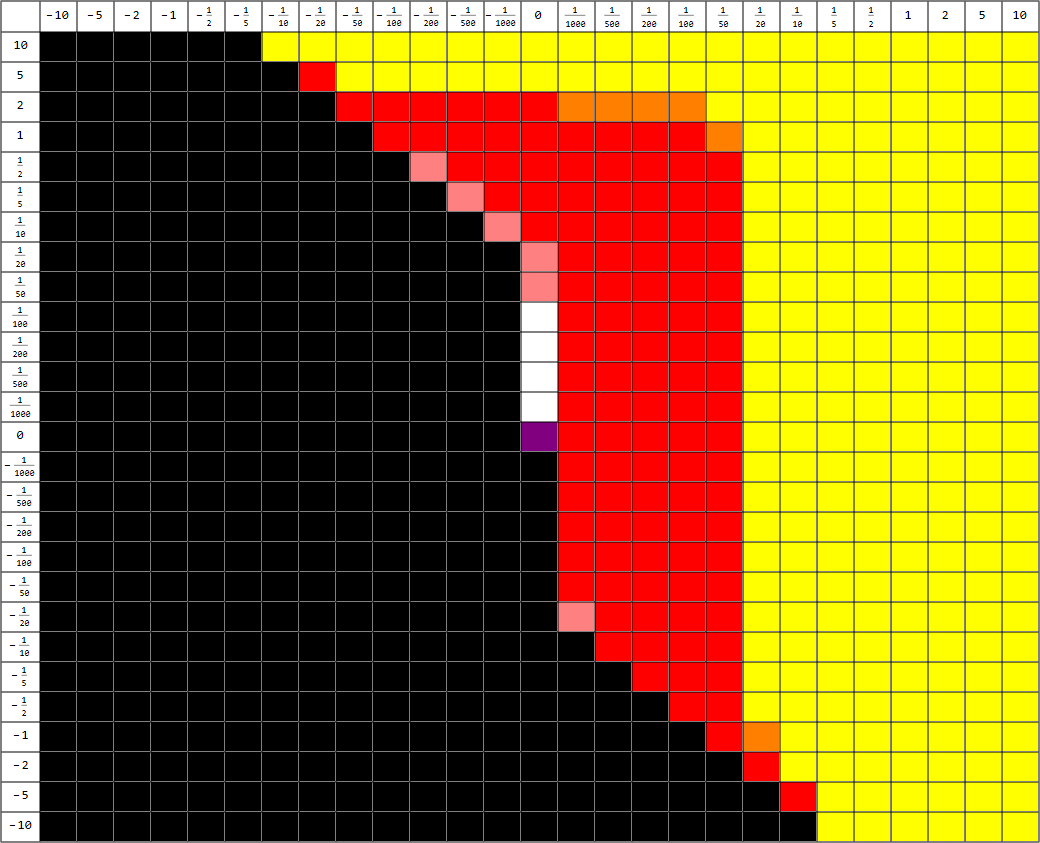} \\ $\beta=1/4$ } & \makecell{\rule{0pt}{29ex}\includegraphics[scale=0.60]{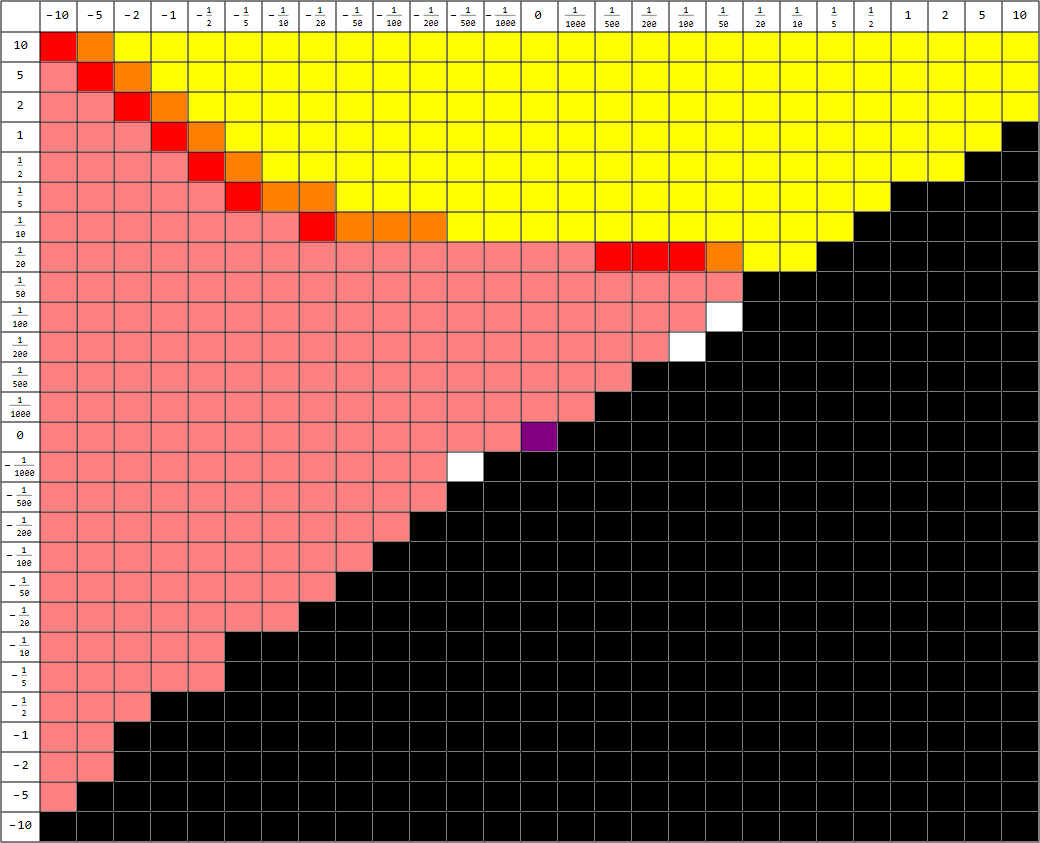} \\ $\beta=1/5$ } \\
		$S_{2}^{-}$ & \makecell{\includegraphics[scale=0.60]{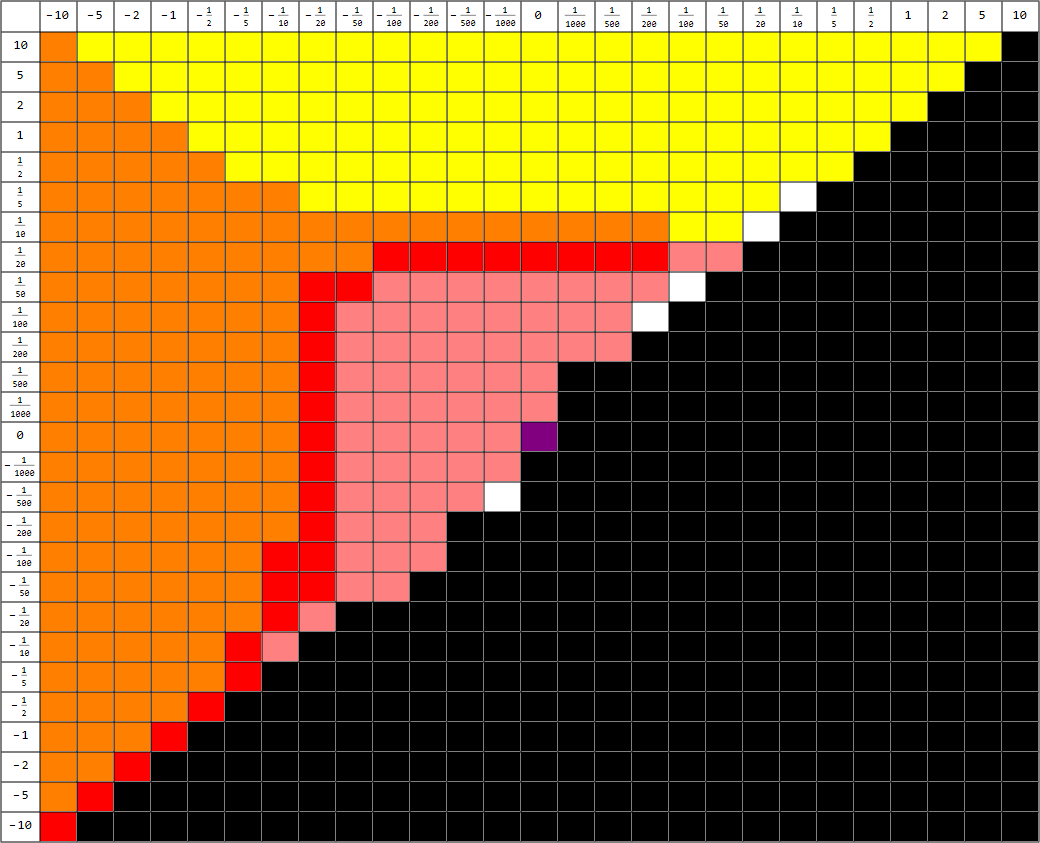} \\ $\beta=1/6$ } & \makecell{\includegraphics[scale=0.60]{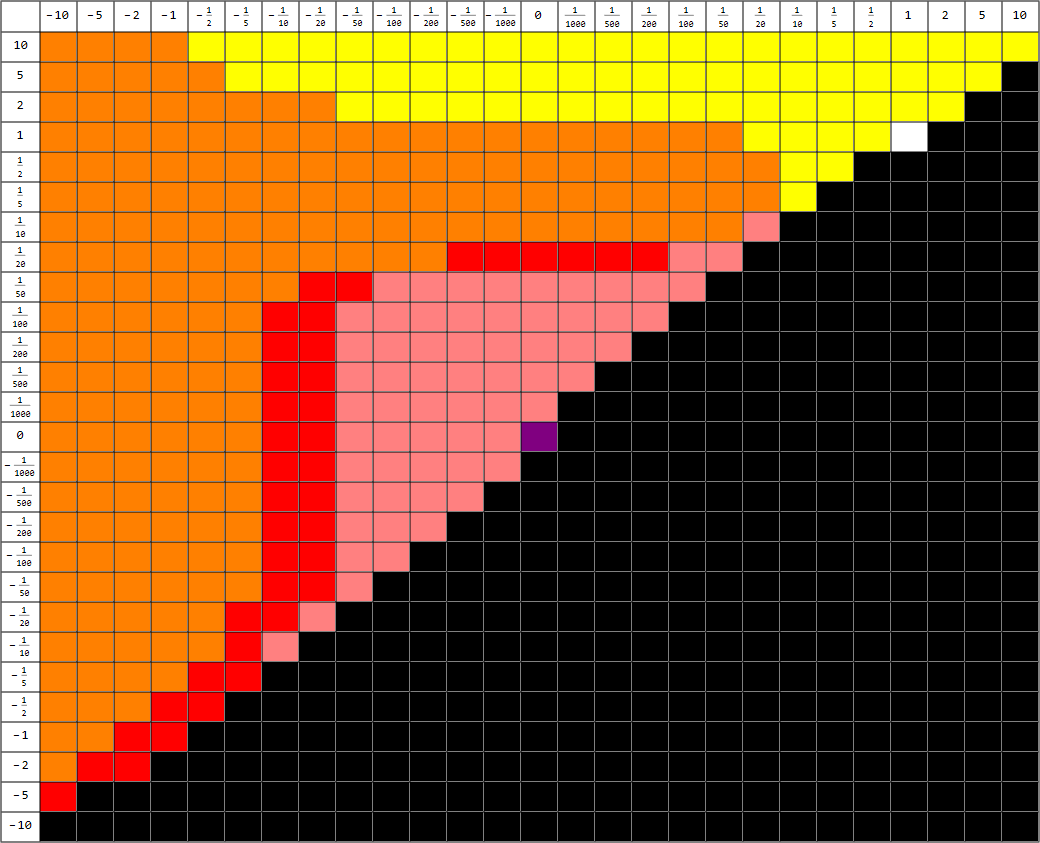} \\ $\beta=13/84$ } & \makecell{\includegraphics[scale=0.60]{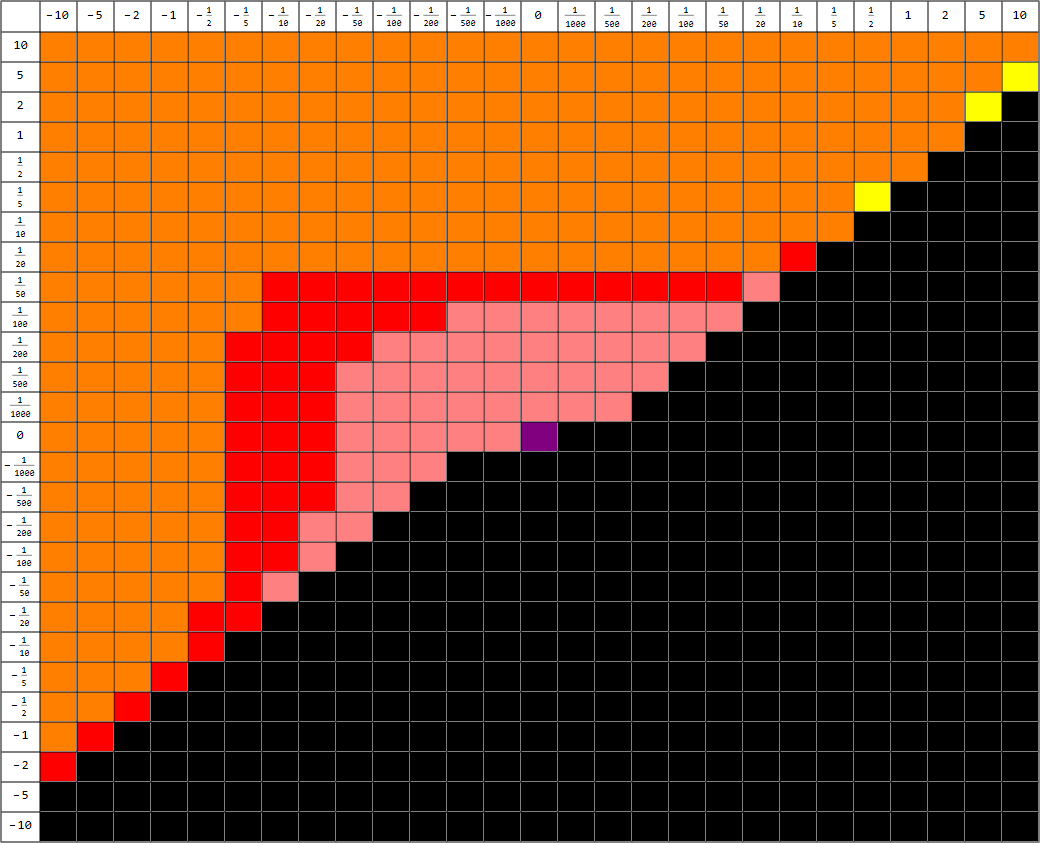} \\ $\beta=1/7$ } \\
		& \makecell{\includegraphics[scale=0.60]{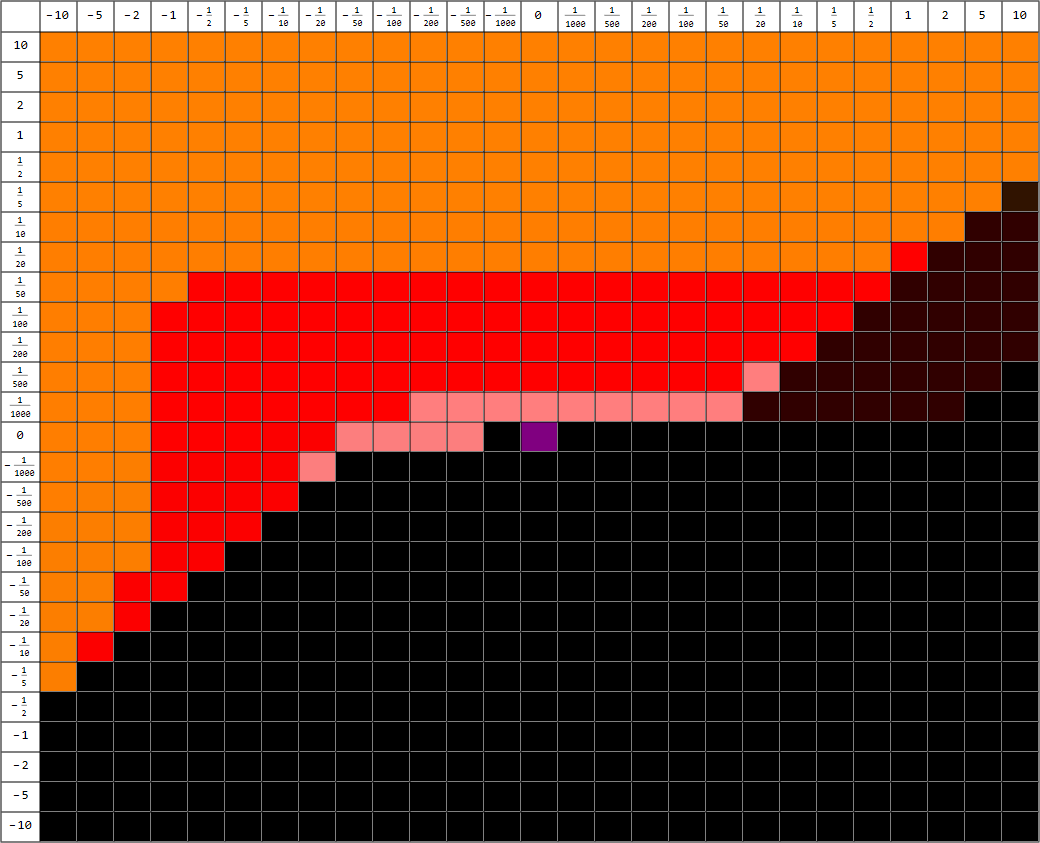} \\ $\beta=1/8$ } & \makecell{\includegraphics[scale=0.60]{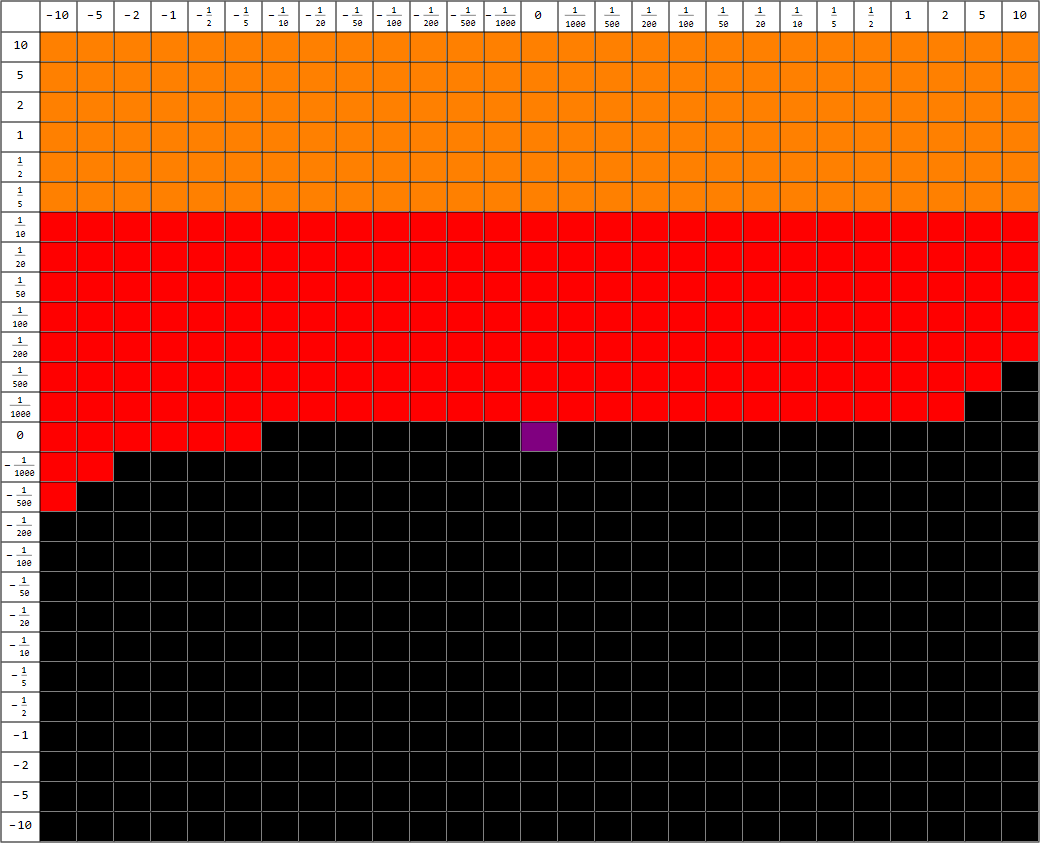} \\ $\beta=1/10$ } & \makecell{\includegraphics[scale=0.60]{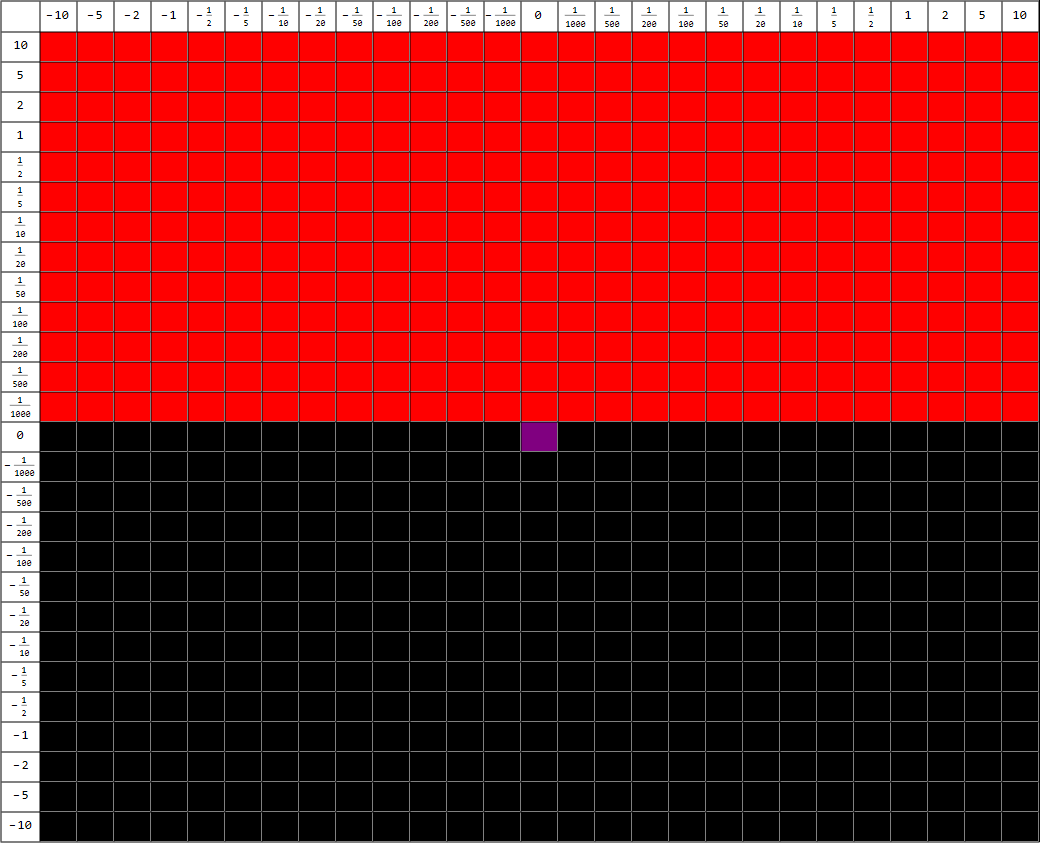} \\ $\beta=1/18$ } \\
		\hline
		
		\multicolumn{4}{|c|}{{\footnotesize Schwarzschild: }\fboxsep=2mm \fboxrule=1px \fcolorbox{black}{purple2}{\phantom{\vrule width 1.5mm height 1.5mm}} \hspace{0.2cm} {\footnotesize Type Ia: }\fboxsep=2mm \fboxrule=1px \fcolorbox{black}{pink2}{\phantom{\vrule width 1.5mm height 1.5mm}} \hspace{0.2cm} {\footnotesize Type Ib: }\fboxsep=2mm \fboxrule=1px \fcolorbox{black}{red}{\phantom{\vrule width 1.5mm height 1.5mm}}  \hspace{0.2cm} {\footnotesize Type Ic: }\fboxsep=2mm \fboxrule=1px \fcolorbox{black}{orange}{\phantom{\vrule width 1.5mm height 1.5mm}} \hspace{0.2cm} {\footnotesize Type II: }\fboxsep=2mm \fboxrule=1px \fcolorbox{black}{yellow2}{\phantom{\vrule width 1.5mm height 1.5mm}} \hspace{0.2cm} {\footnotesize Type III: }\fboxsep=2mm \fboxrule=1px \fcolorbox{black}{black}{\phantom{\vrule width 1.5mm height 1.5mm}} {\footnotesize Num.\ inconclusive: }\fboxsep=2mm \fboxrule=1px \fcolorbox{black}{white}{\phantom{\vrule width 1.5mm height 1.5mm}}
		}\\
		\hline
	\end{tabular}	
	\caption{\label{fig:pspace}
		A number of representative $\beta$-slices through the phase space. The values for $S_0^-$ and $S_2^-$ are $(\pm 10, \pm 5, \pm 2, \pm 1, \pm 0.5, \pm 0.2, \pm 0.1, \pm 0.05, \pm 0.02, \pm 0.01, \pm 0.005, \pm 0.002, \pm 0.001, 0)$.
		Throughout the figures the remaining parameters are fixed at $\gamma=1$, $M=10$, $r_i=35$, and $\alpha=1/2$. Color coding distinguishes the numeric solution classes as indicated in Table \ref{tab:classification}.
	}
\end{figure}

The effective radial potential governing the motion of light-rays within the geometry is given by $V_{\rm eff}(r) = h(r)/r^2$ (see eq.~\eqref{eq:tpeq} below). The scaling properties \eqref{eq:hpowerlaw} then imply that $V_{\rm eff}|_{r=0}$ is finite. This suggests a refined classification based on the value $h_0$, controlling the height of $V_{\rm eff}$ at $r=0$. From the perspective of observations, it is natural to discriminate the three subclasses displayed in Figure \ref{fig:typeIref}:
\begin{description}
	\item [Type Ia] In this case the value of $h_0$ is small in the sense that $V_{\rm eff}(r)$ decreases monotonically for $r < 3M$ while staying positive. In the phase space plot, Fig.\ \ref{fig:pspace}, Type Ia is indicated by pink color.
	\item [Type Ib] Here, $V_{\rm eff}|_{r=0} \leq V_{\rm eff}|_{r=r_{\rm ph}=3M}$. In contrast to Type Ia, the effective potential has a stable minimum between $r=0$ and $r=3M$. In the phase space plot, Figure \ref{fig:pspace}, Type Ib is indicated by red color.
	\item [Type Ic] By definition, these geometries obey $V_{\rm eff}|_{r=0} > V_{\rm eff}|_{r=r_{\rm ph}=3M}$. The effective potential has a stable minimum between $r=0$ and $r=3M$. In the phase space plot, Figure \ref{fig:pspace}, Type Ic is indicated by orange color.
\end{description}

\paragraph{Type II:}
Similarly to Type I, the numerical integration of this class of solutions is reliable up to very small values $r \approx 0$. Again the functions $h$ and $f$ follow the Schwarzschild solution for $r > 2M$ and deviate for $r \le 2M$. The defining property of this class is its characteristic scaling behavior as $r \rightarrow 0$, 
\be
f(r) \sim f_0 \, r^{-1} \, , \quad h(r) \sim h_0 \, r^{-1}  \, . 
\ee
Thus they belong to the analytic class $(-1,-1)_0$ which describes naked singularities \cite{Lu:2015psa}. 
An example solution is illustrated in Figure \ref{fig:extype2}.
In the phase space plot, Figure \ref{fig:pspace}, Type II is indicated by yellow color.
\paragraph{Type III:}
The numerical integration for this solution class terminates at a radius $r_{\rm term}>2M$, at radii slightly larger than the position of the would-be horizon of a Schwarzschild black hole with equal asymptotic mass. Close to the termination point
\be
f(r) \sim f_0 \, (r-r_{\rm term})  \, , \qquad  h(r) \sim h_0 \, ,
\ee
so that these geometries fall into the analytic class $(1,0)_{r_0}$.
This corresponds to wormhole solutions \cite{Lu:2015psa}. An example Type III solution is illustrated in Fig.\ \ref{fig:extype3}. 
In the phase space plot, Fig.\ \ref{fig:pspace}, Type III is indicated by black color.

With the classification of the individual phase space elements completed, we now proceed to construct the phase space of static, asymptotically flat, and spherically symmetric vacuum solutions in quadratic gravity. Since the solution space is spanned by five free parameters, we reduce the complexity of the problem by fixing the asymptotic mass of the solutions, $M=10$ in natural units\footnote{Irrespective of the concrete choice, the modifications of the solutions are of topological nature and therefore remain in an astrophysical setting.}, and the mass of the massive spin-two degree of freedom, $m_2 = 1$ (equivalently, $\alpha = 1/2$). Since $M$ is accessible via observations in the weak-gravity regime analyzing the solution space keeping this variable fixed is natural. At the same time, keeping $\alpha$ fixed and varying $\beta$ allows to investigate the relative impact of the  quadratic gravity contributions associated with the massive spin-two and spin-zero degrees of freedom. The remaining three parameters, $\beta$, $S_0^-$, and $S_2^-$, are varied to construct representative 2-dimensional slices of the phase space. For our purposes, $\beta$ takes the values $(1/3$, $1/4$, $1/5$, $1/6$, $13/84$, $1/7$, $1/8$, $1/10$, $1/18)$. For each value of $\beta$, we investigate all combinations of the Yukawa amplitudes $S_0^-$ and $S_2^-$ taking values in $(\pm 10$, $\pm 5$, $\pm 2$, $\pm 1$, $\pm 0.5$, $\pm 0.2$, $\pm 0.1$, $\pm 0.05$, $\pm 0.02$, $\pm 0.01$, $\pm 0.005$, $\pm 0.002$, $\pm 0.001$, $0)$. In total, this results in approximately $10^4$ spacetime geometries.

We then use \eqref{eq:linsolflat} to impose initial conditions at $r_i = 3.5 M$ and integrate inwards, identifying the corresponding solution class by matching the numerical result to the analytic scaling behavior close to the termination point. The distribution of the numeric solution classes are visualized in Fig.\ \ref{fig:pspace} with the color code in Table \ref{tab:classification}. This clearly illustrates that quadratic gravity admits a rich phase space of black-hole type solutions. Notably, the Schwarzschild solution appears for all values $\alpha$, $\beta$ and is situated at	$(S_0^-,S_2^-) = (0,0)$. It constitutes the only solution with an event horizon. This is in agreement with the expectation that there is no infinitesimal deformation of Schwarzschild in the context of quadratic gravity which allows a smooth transition to another solution class \cite{Lu:2015psa}. 

At this point it is worthwhile to highlight several features of our result. Let us start with the case $\beta = 1/6$ which corresponds to the spin-zero and spin-two mass being equal, $m_2=m_0=1$. Then both Yukawa potentials show the same damping and we can realize all types of geometries by dialing $S_2^-$ and $S_0^-$. Decreasing $\beta$ leads to an increase of $m_0$. This results in a relative suppression of the spin-zero contributions with respect to the spin-two part. As a consequence the initial conditions become essentially independent of  $S_0^-$ so that the lines separating different phases turn vertically (c.f.\ the subfigure for $\beta = 1/3$). Conversely, increasing $\beta$ beyond the equal mass case leads to a relative suppression of the spin-two contribution and the phase space becomes essentially independent of $S_2^-$. This effect underlies the horizontal stripes obtained at $\beta = 1/18$. 

\begin{table}[t!]
\centering
\begin{tabular}{|cccl|}
\hline
Solution class & $\Big.$ Analytic class & Color in Fig.~\ref{fig:pspace} & Defining characteristic \\ \hline \hline
Type S & $\Big. (1,1)_{r_0}$ & purple & Schwarzschild geometry (unique) \\
\hdashline
Type Ia & $\Big. (-2,2)_0$ & pink & naked singularity, no stable photon orbit below $r = 2M$  \\
Type Ib & $\Big. (-2,2)_0$ & red & naked singularity, stable photon orbit below $r = 2M$\\
Type Ic & $\Big. (-2,2)_0$ & orange & naked singularity, partially screened by a reflective barrier\\
\hdashline
Type II & $\Big. (-1,-1)_0$ & yellow & \makecell[cl]{naked singularity, fully screened by a reflective barrier} \\
\hdashline
Type III & $\Big. (1,0)_{r_0}$ & black & wormhole throat at finite $r_{\rm term} > 2M$ \\
\hline
None & None & $\Big.$ white & inconclusive initial conditions due to numerical instability \\
\hline
\end{tabular}
\caption{\label{tab:classification}
Mapping numeric solution classes to analytic classes. Characteristic properties of the asymptotically flat vacuum solutions in quadratic gravity are presented, see Fig.~\ref{fig:pspace}.}
\end{table}

We checked the robustness of the phase space classification shown in Fig.\ \ref{fig:pspace} by varying the value $r_i$ where the initial conditions are imposed. While this resulted in minor changes in the functions $h(r)$ and $f(r)$, the analytic scaling class of the solutions remained largely unaffected. The small number of cases where the change of $r_i$ actually triggered a change in the classification are indicated as white squares. Notably, these appear close to a phase transition line, indicating that our numerical construction is not sufficiently elaborate to associate the point to a specific phase.

An intriguing consequence of the phase space shown in Fig.\ \ref{fig:pspace} is that it is actually quite difficult to discriminate general relativity and quadratic gravity based on its static, spherically symmetric, and asymptotically flat solutions. Since the Schwarzschild geometry solves both equations of motion any experiment confirming the validity of this solution can be understood as an experimental verification of both theories. Distinguishing among the two settings is possible if and only if an experiment actually detects a deviation from the GR, which could then point towards an alternative geometry appearing in the phase space of quadratic gravity.

Notably, the stability of spacetimes with potentials similar to the ones encountered for the solutions Type Ic and Type II (and potentially also Type Ib) have been discussed in the context of ultracompact objects in \cite{Cardoso:2014sna,Keir:2014oka}. In this context it was argued that gravitational wave perturbations could potentially destabilize the geometry. We refrain a detailed discussion of such an effect for our our geometries to future work.

\section{Constraining Quadratic Gravity through its Shadow}
\label{sec.3}
In the previous section, we demonstrated that the phase space of static, spherically symmetric and asymptotically flat solutions appearing in quadratic gravity comes with two sets of free parameters. Firstly, the masses $m_2$ and $m_0$ introduced in eq.\ \eqref{eq:masses} are directly related to the coupling constants associated with the higher-derivative interactions. Secondly, the parameters $\left( M, S_2^-, S_0^- \right)$, with $M$ being the asymptotic mass of the solution, parameterize the solution space for fixed masses $\left(m_2, m_0\right)$. In order to constrain the additional parameters for quadratic gravity through observations, we systematically compute observables from the underlying geometries. Since the solutions spanning the phase space are virtually degenerate far away from the gravitational source, the focus is on the strong gravity regime situated within the unstable photon orbit of the geometries. Practically, we limit ourselves to observables, which are accessible by the Event Horizon Telescope. The coordinate radius of the outer (unstable) photon orbit is considered in Sect.\ \ref{sec.31}. Here we demonstrate that this observable turns out to be virtually identical for the Schwarzschild solution and the geometries associated with naked singularities. Following \cite{Bambi:2013nla,Shaikh:2018lcc}, we then consider the intensity profile of light geodesics originating from radially freely falling spherically symmetric accreating matter around the gravitational source. Throughout this section, we work in the probe approximation, neglecting matter-self interactions as well as the backreaction of the matter on the geometry.

\subsection{Photon Rings}
\label{sec.31}
Making the implicit assumption that the radius separating the plunge rays from the rays going out to infinity coincides with the position of the unstable photon sphere, Ref.\ \cite{EventHorizonTelescope:2020qrl} derived that, for spherically symmetric spacetimes, the coordinate radius $r_{\rm sh}$ of the shadow measured by a distant observer depends on the $tt$-component of the metric. Based on the areal coordinate system used in \eqref{eq:metans} one finds
\be\label{eq:shadow}
\rsh = \frac{\rph}{\sqrt{h(\rph)}} \, .
\ee
Here $\rph$ is the coordinate radius of the unstable photon orbit, determined from the condition
\be\label{eq:rph}
\rph = 2 \, h(\rph) \left( \left. \frac{d h(r)}{dr} \right|_{r = \rph} \right)^{-1} \, . 
\ee
The angular radius of the photon ring on the sky is then given by $\theta_{\rm ph} = \rph/D$ with $D$ the distance to the object. In practise, observations give $\theta_{\rm ph} = 42 \pm 3$ $\mu$as \cite{EventHorizonTelescope:2019PaperI}.

For the Schwarzschild case \eqref{eq:ssmet} one recovers $\rph = 3M$ and the coordinate radius of the shadow evaluates to
\be
\rsh^{\rm Schwarzschild} = 3 \sqrt{3} \, M \, . 
\ee
It is then suggestive to define
\be
\zeta \equiv \frac{\rsh^{\rm model} - \rsh^{\rm Schwarzschild}}{\rsh^{\rm Schwarzschild}} 
\ee 
which measures \emph{the deviation of the shadow position} as compared to the Schwarzschild case. \\

We now turn our attention to the photon ring radius assuming the linearized solution in \eqref{eq:linsolflat} still holds, with the goal of evaluating the deviation parameter $\zeta$ for the astrophysical black hole M87$^\ast$ in mind. Evaluating equation \eqref{eq:rph} for the linearized solution we obtain, in linear approximation
\begin{equation}
    r_{\rm ph} = 3\,M + 3\,S_2^-\,(m_2\,M - 1)\,e^{-3\,m_2\,M} + \frac{3}{2}\,S_0^-\,(m_0\,M - 1)\,e^{-3\,m_0\,M}.
\end{equation}
For this formula to be valid one needs the condition $3\,m_i\,M \gg 1$ to be satisfied for both relevant masses $m_0, m_2$. If the linearized solution is valid near the photon ring, it is also valid outside of it and therefore the linearized solution can also be plugged into equation \eqref{eq:shadow} to obtain $r_{\rm sh}$. Setting, for simplicity, $S_0^- = 0$ (or equivalently for the purposes here taking $\beta = 0$), we have for the shadow radius

\begin{equation}
    r_{\rm sh} = 3 \sqrt{3} \, M \, + 3 \sqrt{3} \, S_2^- \, (2\,m_2\,M + 1)\, e^{-3\,m_2\,M}.
\end{equation}
Note that generally the smaller $m_2$ the bigger the deviation of the shadow radius is compared to the Schwarzschild case. Assuming $m_2$ to be of the order of the Planck mass, say $m_2 = m_{\rm pl}$, and taking for $M = 2.2 \times 10^{50} \, m_{\rm pl}$, roughly the mass of the black hole M87$^\ast$, we get for the deviation parameter
\begin{equation}
    \zeta^{\rm M87} = 2 \, \rm exp\left(-\,6.5 \times 10^{50}\right)\, S_2^-
\end{equation}
Clearly, for any seemingly realistic values of $S_2^-$, this number is zero for all practical purposes. This serves as a typical example to how small quantum gravity corrections can be, and serves as a direct consequence that the modifications of the quadratic gravity terms do not (significantly) extend to the photon radius. However, since the modifications do extend to just beyond the would-be horizon we will now discuss another observable that \emph{can} be seriously affected by the solutions presented in sect.\ \ref{sec.2}.

\subsection{Intensity Profiles from Accreting Matter}
A second observable related to observations by the Event Horizon Telescope is the intensity profile of radiation emitted by surrounding matter. In this work, we study this observable based on the optically thin accreting matter model proposed in \cite{Bambi:2013nla}.\footnote{This model has already been used to compare the intensity profiles of a Schwarzschild black hole and a worm hole geometry \cite{Bambi:2013nla} and subsequently, Schwarzschild black holes and naked singularities given by the Joshi-Malafarina-Narayan spacetimes \cite{Shaikh:2018lcc,Kaur:2021wgy}, showing that the difference in geometry leads to distinguished features in the intensity profile.} The key advantage of this model is that it is sufficiently simple so that the intensity profiles can be obtained without numerically expensive ray tracing techniques. In this way, one can actually evaluate the profiles for a large number of geometries comprising the phase space discussed in Sect.\ \ref{sec.23}.\footnote{For the proposal that the emission comes from the jet of astrophysical black holes, esp.~in the case of Sgr A$^\ast$ and M87$^\ast$, see \cite{Falcke:1993kd,Falcke:2000nai}. Accretion flows have been reviewed in \cite{Yuan:2014gma}.}

The matter, accreting to the center of gravity, is taken to be a spherically symmetric, optically thin, and radially freely falling. It emits monochromatic radiation whose emissivity in the emitter frame per unit volume falls of proportional to $r^{-2}$, see also \cite{2000}:
\be
j(\nu_e) \propto \frac{\delta(\nu_e - \nu_*)}{r^2} \, . 
\ee
Here $\nu_*$ is the emitted photon frequency in the rest frame of the emitter. We are then interested in the intensity of light detected by an observer at rest situated at a distance $D \gg M$ from the black hole. Introducing $X,Y$ as the coordinates on the asymptotic observer's screen, the intensity of photons observed at frequency $\nu_{\rm obs}$ is given by \cite{Jaroszynski:1997bw}
\be\label{Iobs}
I_{\rm obs}(\nu_{\rm obs}, X,Y) = \int_\gamma g^3 \, j(\nu_e) \, dl_{\rm prop} \, . 
\ee
Here $\gamma$ denotes the path taken by the photon, $g = \nu_{\rm obs}/\nu_e$ is the redshift factor, and $dl_{\rm prop}$ is the infinitesimal proper length measured in the rest frame of the emitter. These can be expressed in terms of the 4-momentum of the photon $k^\mu$, the 4-velocity of the observer $u^\mu_{\rm obs} = (1,0,0,0)$ and the 4-velocity of the accreting matter $u_e^\mu$. Starting from \eqref{eq:metans}, and considering a spherically symmetric gas in radial free-fall the later is given by
\be
u^t_e = \frac{1}{h} \, , \qquad u^r_e = - \sqrt{\frac{(1-h) f}{h}} \, , \qquad u^\theta_e = u^\phi_e = 0 \, . 
\ee
For a photon the $t$-component of the $4$-momentum $k_\mu$ is a constant of integration. For radial motion, the $r$-component can then be obtained from the fact that $k^\mu$ is light-like, satisfying $k_\mu k^\mu = 0$. Using the Euler-Lagrange equations to infer that $k_t = E$ and $k_\phi = L$ are constants of motion, one has
\be\label{eq:4momentum}
k_r = \pm k_t \, h \, \sqrt{\left( \frac{1}{h} - \frac{b^2}{r^2} \right)\,f\, } \, , 
\ee
with $b^2 \equiv L^2/E^2$ being the impact parameter of the ray. The $+$($-$)-sign holds when the photon approaches (goes away from) the center of gravity. The redshift factor is given by
\be
g = \frac{k_\alpha u^\alpha_{\rm obs}}{k_\beta u^\beta_{e}} \, . 
\ee 
Substituting the explicit expressions for the 4-velocities together with \eqref{eq:4momentum}, then gives
\be\label{eq:redshift}
g_{\pm} = \left( \frac{1}{h} \mp \frac{|k_r|}{k_t} \sqrt{(1-h)\frac{f}{h}\, } \right)^{-1} \, , 
\ee
where the sign again refers to photons going towards (or away from) the center of gravity. Finally, the proper distance $dl_{\rm prop} = k_\alpha u^\alpha_e d\lambda$ can be evaluated in terms of the photon's 4-momentum and the redshift factor
\be
dl_{\rm prop} = \frac{k_t}{g |k^r|} dr \, . 
\ee
Substituting these intermediate results into \eqref{Iobs} and integrating over all frequencies finally gives \cite{Bambi:2013nla}
\be\label{Iobsfinal}
I_{\rm obs}(X,Y) \propto \int_\gamma \frac{g^3 \, k_t \, dr}{r^2 |k^r|} \, . 
\ee
For an observer sufficiently far away from the center of gravity, $X^2 + Y^2 = b^2$ and the right-hand side of \eqref{Iobsfinal} depends on the impact parameter $b$ only. Owed to the spherical symmetry of the model, it is then sufficient to calculate the intensity as a function of $b$ only.

\begin{figure}[t!]
	\centering
	\begin{subfigure}{0.3\textwidth}
		\centering
		\includegraphics[width=\textwidth]{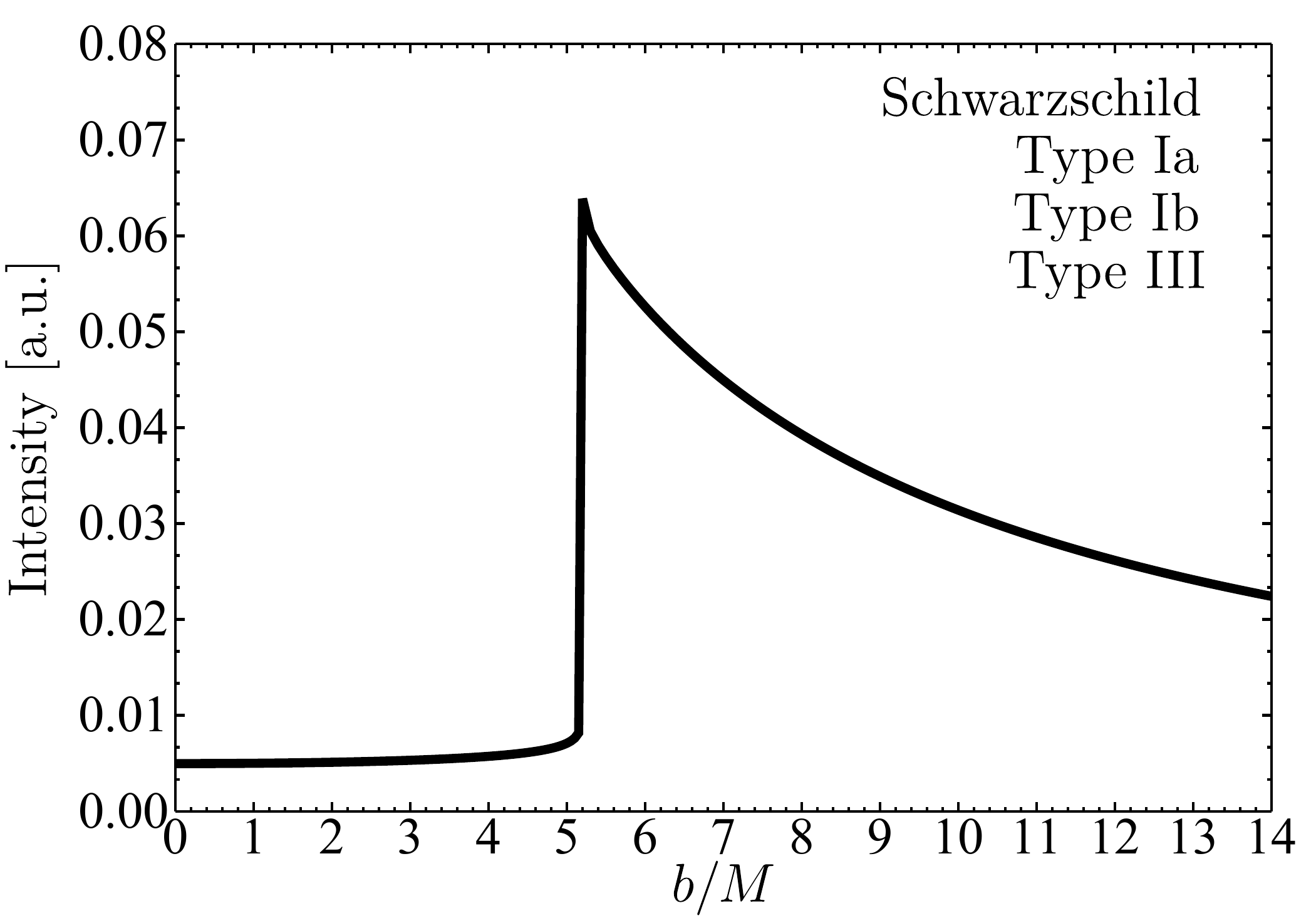}\\[3ex]
		\includegraphics[width=\textwidth]{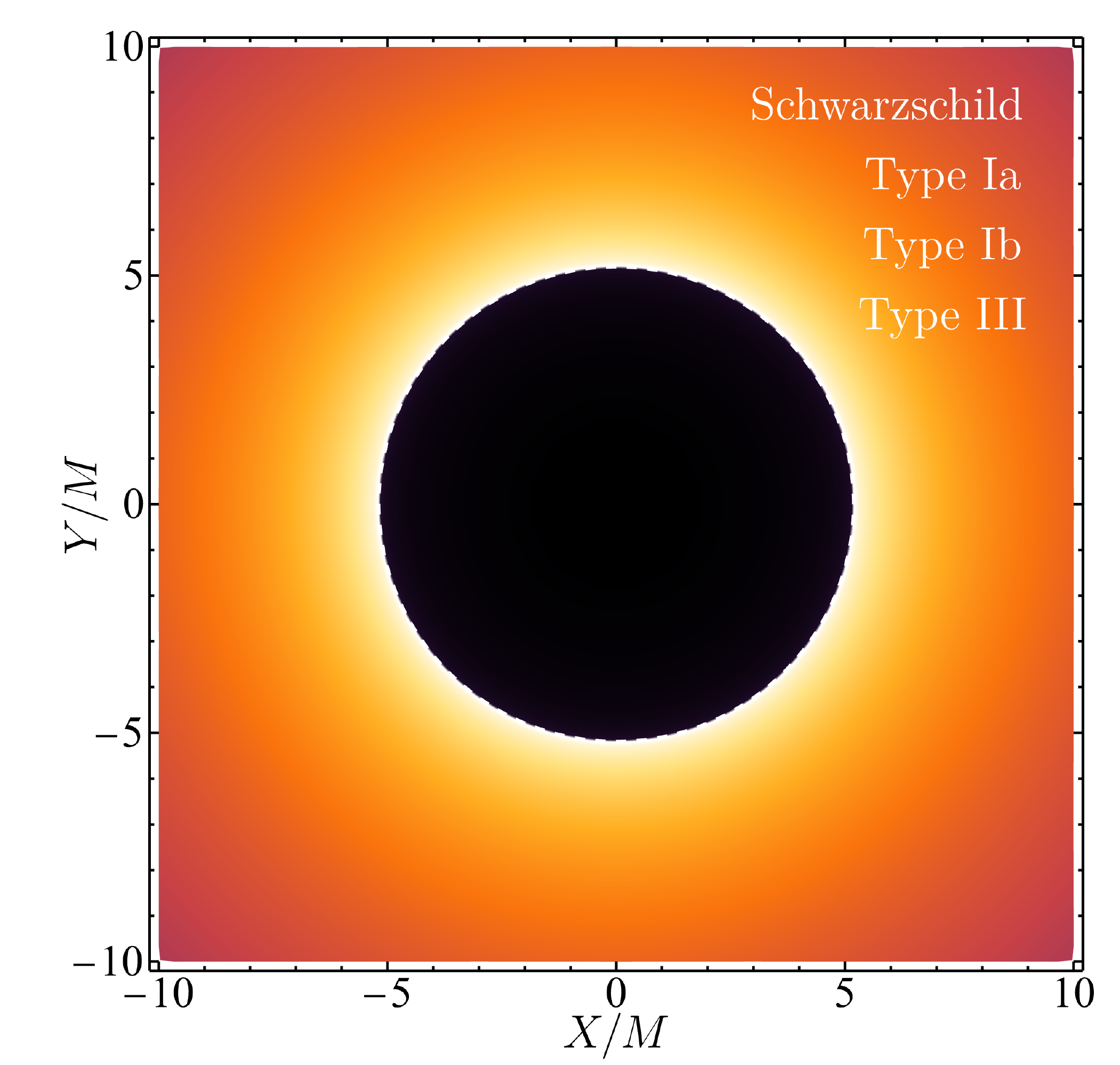}
		\caption{\label{fig:intss} Schwarzschild}
	\end{subfigure}
	\hfill
	\begin{subfigure}{0.3\textwidth}
		\centering
		\includegraphics[width=\textwidth]{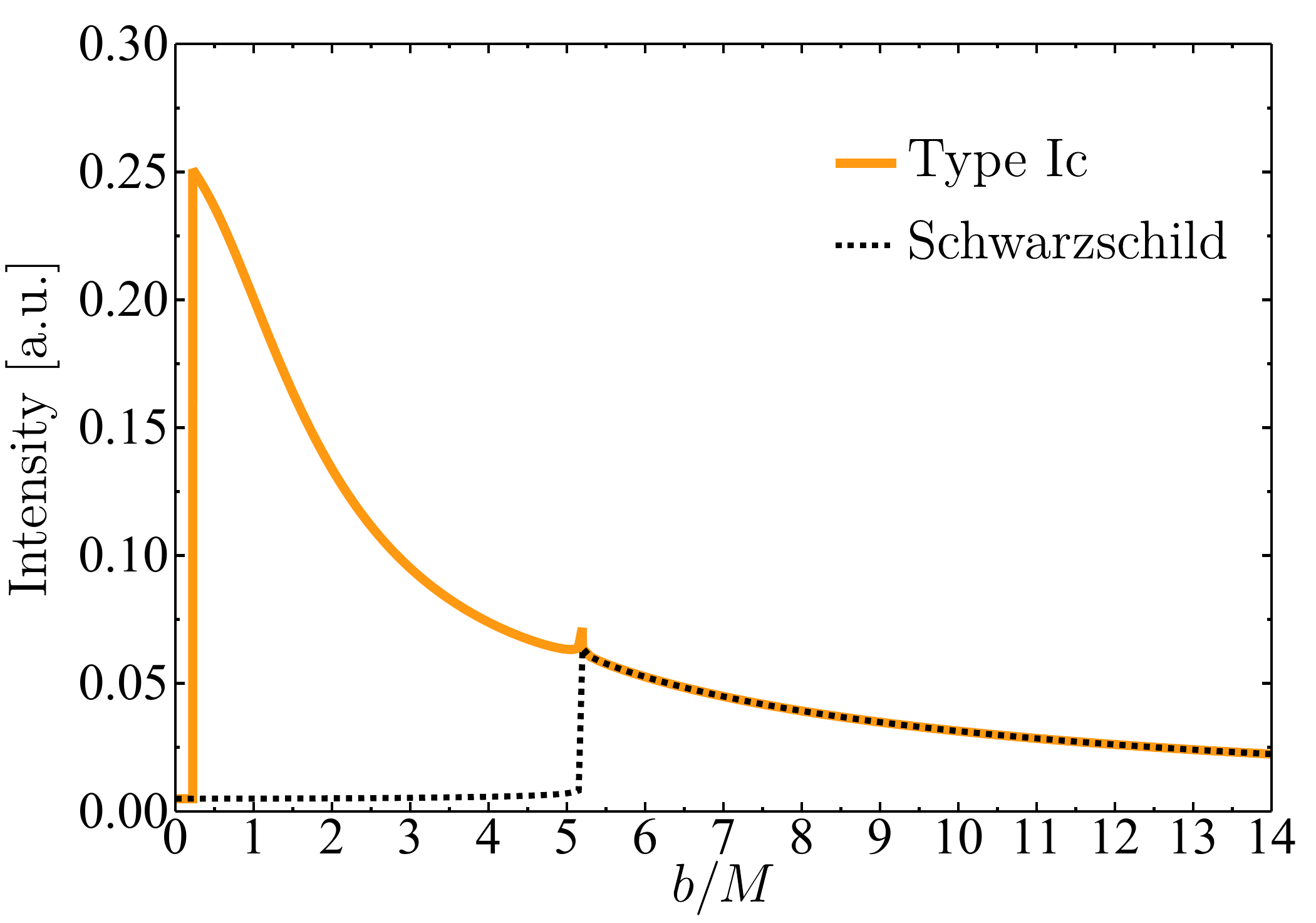}\\[3ex]
		\includegraphics[width=\textwidth]{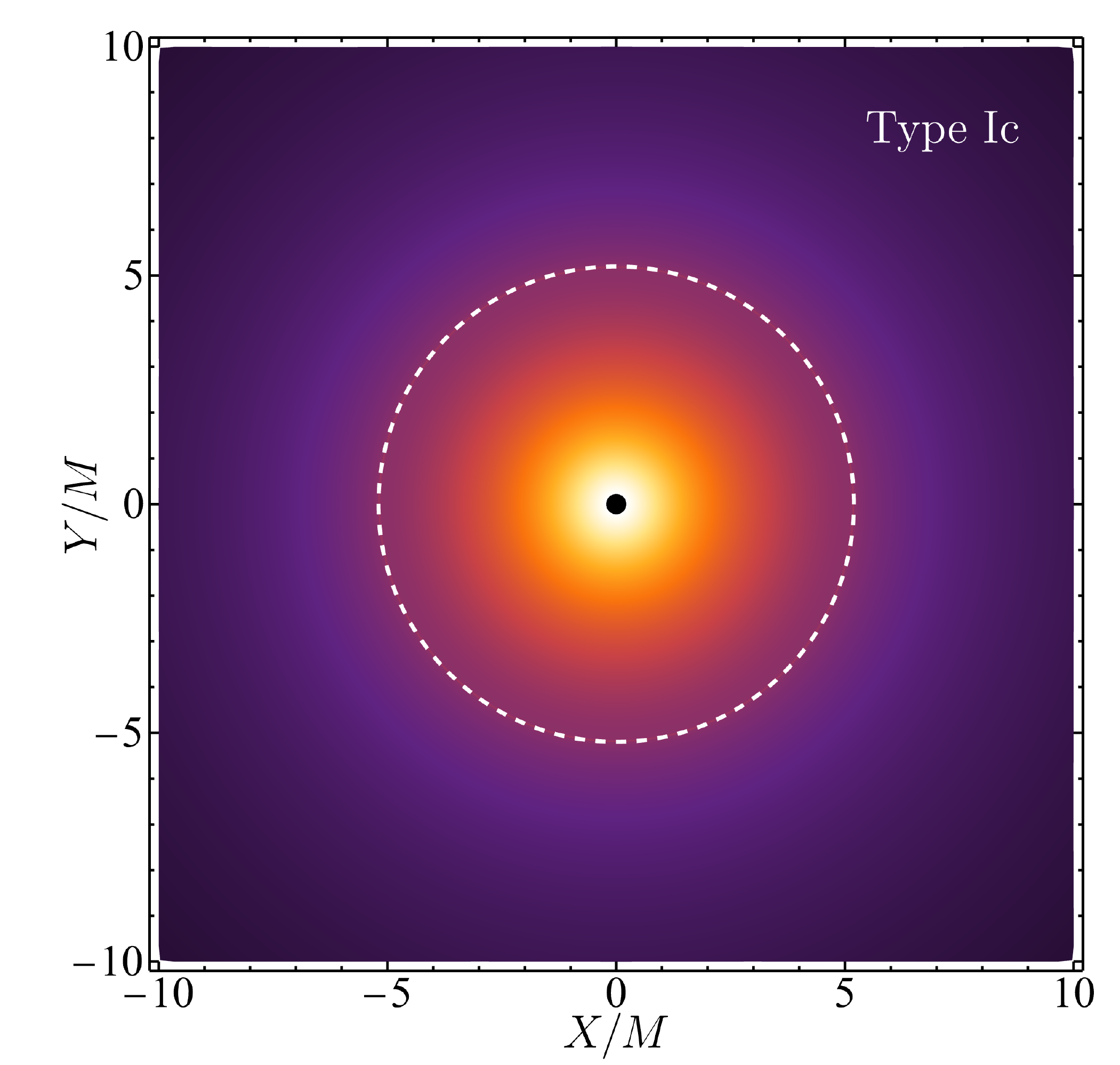}
		\caption{\label{fig:inttype1c}Type Ic}
	\end{subfigure}
	\hfill
	\begin{subfigure}{0.3\textwidth}
		\centering
		\includegraphics[width=\textwidth]{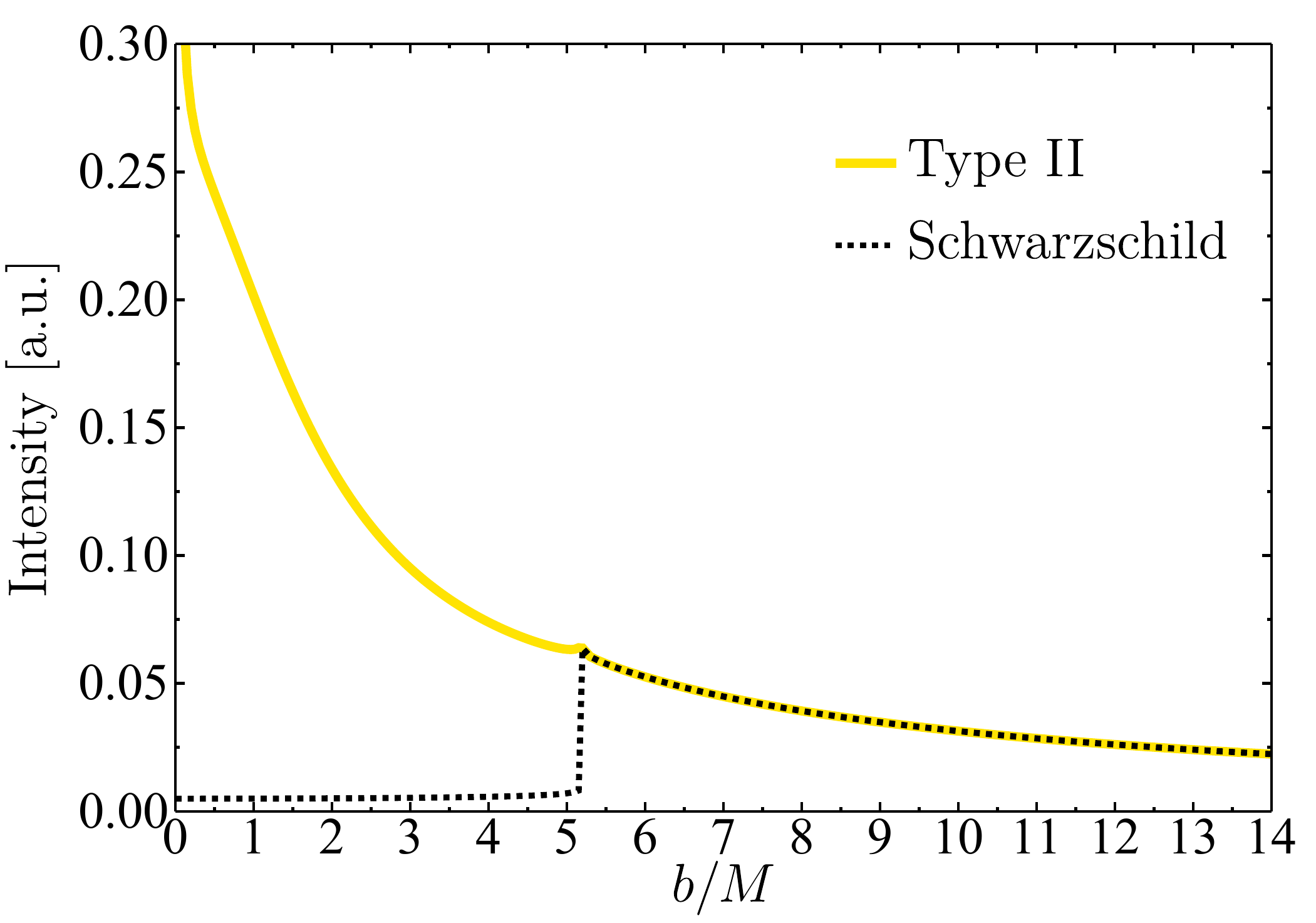}\\[3ex]
		\includegraphics[width=\textwidth]{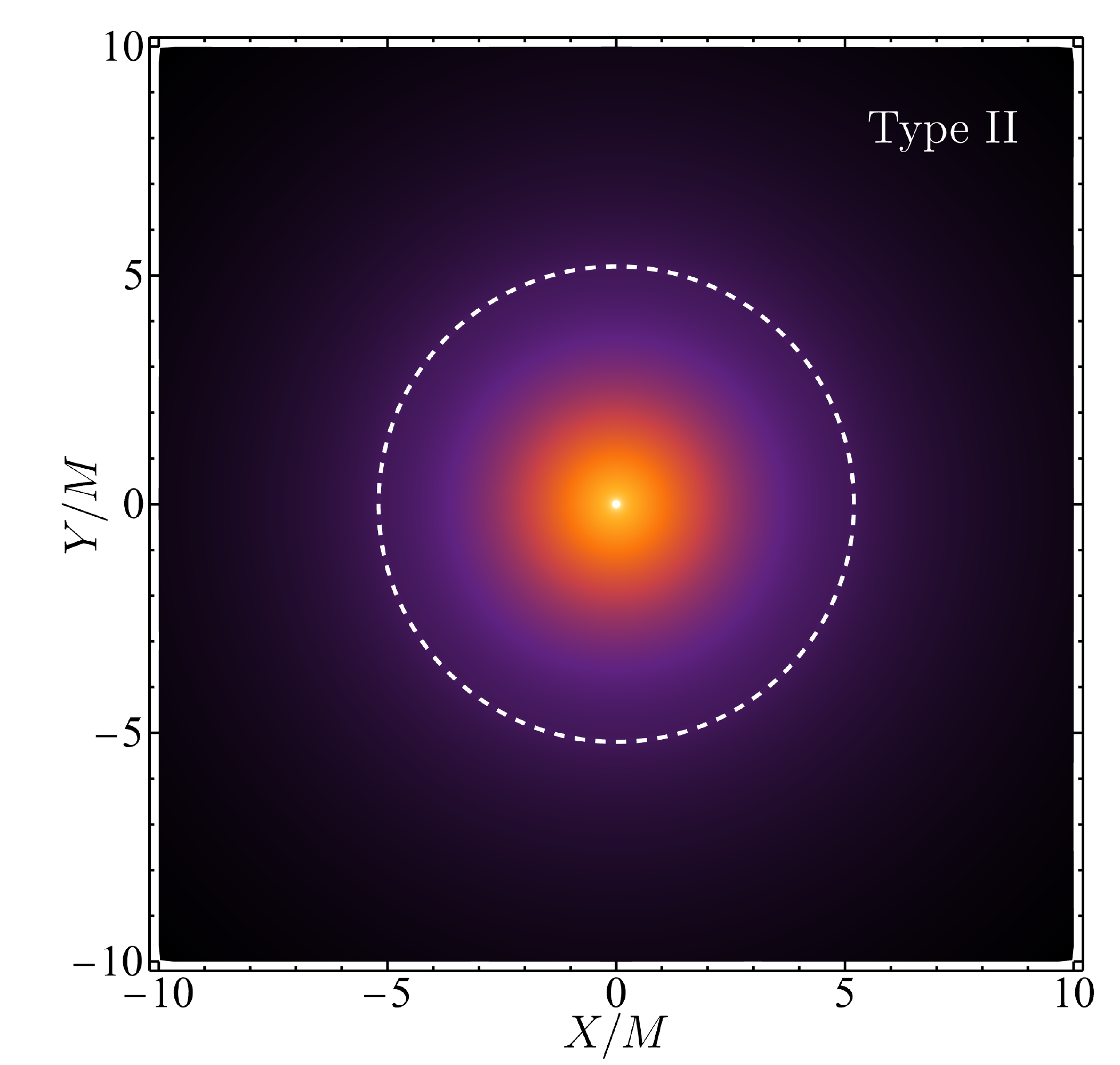}
		\caption{\label{fig:inttype2}Type II}
	\end{subfigure}
	\caption{\label{fig.intense} Distinguished intensity profiles obtained from for the emission model \eqref{Iobsfinal} for spacetimes obtained from the initial conditions listed in Table \ref{tab:initialconditions}. The left column shows the intensity for a Schwarzschild black hole with $M=10$ (top row). This intensity profile is virtually indistinguishable from naked singularities of Type Ia, Type Ib, and Type III. The middle column gives the intensity for the Type Ic solution with initial conditions. The size of the shadow of a Schwarzschild black hole is added as a dashed line for reference. The presence of an inner turning point significantly enhances the intensity in the region $b_{\rm tp2} \le b \le b_{\rm crit}$. Intensity profiles for Type II solutions illustrated in the right column do not exhibit a shadow; their image resembles the one of a star with the peak intensity at small impact parameter $b$.}
\end{figure}
In practice, we construct the intensity profiles as follows. The screen, corresponding to the telescope situated far away from the gravitational source, is placed at a coordinate distance $d_s = 10^5 M$ away from the center of gravity. Starting from the screen, the light rays are then traced backwards through spacetime. Owed to the spherical symmetry of the geometry, it suffices to trace rays within the equatorial plane $\theta = \pi/2$. For these paths
\be\label{eq:conserved}
\frac{dt}{ds} = \frac{E}{h(r)} \, , \qquad \frac{d\phi}{d s} \equiv \frac{L}{r^2}
\ee
are integrals of motion. This allows to convert the condition for light-like geodesics, $g_{\mu\nu} \frac{dx^\mu}{ds} \frac{dx^\nu}{ds} = 0$, into a first order equation determining $r(s)$
\be\label{eq:tpeq}
\left( \frac{dr}{ds} \right)^2 = \frac{f(r)}{h(r)} \, L^2 \, \left( \frac{1}{b^2} -  V_{\rm eff}(r)\right) \, ,  
\ee
where $V_{\rm eff}(r) \equiv h(r)/r^2$.
Based on the explicit value for $b$ one distinguishes two types of light rays. If $\frac{dr}{ds} = 0$ at some radius $r$, the photon trajectory $\gamma$  has a turning point there and continues outward towards infinity. In this case, the total intensity picked up along the ray consists of two contributions: for the part of $\gamma$ connecting the screen to the turning point, eq.\ \eqref{Iobsfinal} is evaluated choosing the minus-sign in \eqref{eq:redshift} while the intensity picked up on the other side of the turning point is given by \eqref{Iobsfinal} evaluated with the redshift factor for photons going towards the source. The observed intensity is then given by the sum of these contributions. The second class of light rays does not posses a turning point and therefore goes on until it reaches a termination point. For the Schwarzschild geometry the rays terminate at the event horizon while for geometries of Type I the rays reach $r=0$ where they hit the curvature singularity. Now, $r(s)$ decreases monotonically and the intensity formula is integrated from the termination point of the object to the screen, using the minus sign in \eqref{eq:redshift}.

At this stage, several important remarks on evaluating \eqref{Iobsfinal} are in order. For $r > 3M$ all our example geometries agree with the Schwarzschild geometry up to exponentially suppressed correction terms. This feature entails that light rays not entering the region $r < 3M$ see the same geometry irrespective of the solution. The tail of the intensity profiles where $b > b_{\rm crit}$ is universal and shared by all geometries: this part of the profile just probes spacetime in regions where the quadratic gravity effects are negligible. Thus the evaluation can be speed up by just focusing on light rays with $b < b_{\rm crit}$ and completing the intensity profile with the Schwarzschild result for $b > b_{\rm crit}$. Secondly, this feature allows to bypass the need of extending the numerically constructed solutions up to the observer screen which owed to the instability of the equations would result in a insurmountable fine-tuning problem for the initial conditions. Instead the outer region of spacetime can just be taken to be the Schwarzschild solution with the gluing point taken at $r=3M$. Finally, evaluating the intensity profile for wormhole geometries (Type III) requires additional assumptions. In this case $f(r_t) = 0$ at a finite value $r_t > 2M$ and the geometry can be extended to $r < r_t$ in such a way that $r_t$ describes the throat of the wormhole. We then adopt the assumption made in \cite{Bambi:2013nla} that no radiation comes out of the hole. Thus the intensity picked up along rays connecting to the throat is obtained by integrating \eqref{Iobsfinal} from $r_t$ to the screen. 

The characteristic intensity profiles as a function of the ray's impact parameter $b$ associated with the geometries generated from the initial conditions in Table \ref{tab:initialconditions} are shown in Fig.\ \ref{fig.intense}. The reference profile obtained from the Schwarzschild geometry is depicted in the left column. In this case the intensity shows a step rise at $b = b_{\rm crit} = 3 \sqrt{3}M$, marking the impact parameter for which the ray touches the unstable photon orbit. For $b < b_{\rm crit}$ the rays plunge into the black hole horizon, creating the low-intensity region associated with the black hole shadow shown in the lower-left figure. The intensity profiles resulting from the geometries of Type Ia, Type Ib, and Type III are indistinguishable from the reference plot and are thus not shown separately. Thus, despite the region $0 < r < 2M$ contributing to the intensity profiles of Type Ia and Type Ib, it results in a negligible contribution.\footnote{For Type Ib, a more sophisticated emission model may bypass this conclusion owed to the stable minimum exhibited by $V_{\rm eff}$. Detailed studies based on the RAPTOR ray-tracing code \cite{Bronzwaer:2018lde,Bronzwaer:2020kle} are currently ongoing and will be reported elsewhere \cite{inprep}.}  Similarly, the regions where the Type III geometry differs from the Schwarzschild reference also give negligible contributions to the intensity profile.

The intensity profile generated by geometries of Type Ic is shown in the middle column of Fig.\ \ref{fig.intense}. These geometries again exhibit a shadow region, albeit with smaller radius $b_{\rm Type-Ic-shadow} < b_{\rm crit}$. This effect is shown in the lower-middle figure where the shadow size of the Schwarzschild black hole with identical asymptotic mass is superimposed as the white dashed circle. The interval $b_{\rm Type-Ic-shadow} < b < b_{\rm crit}$ is actually brighter than the intensity found in the tail at $b > b_{\rm crit}$. This is a direct consequence of the characteristic property of Type Ic geometries where $V_{\rm eff}|_{r=0} > V_{\rm eff}|_{r = 3M}$ by definition, see Fig.\ \ref{fig:typeIref}. Owed to this property, light rays entering the region $r < 3M$ can still exhibit a turning point and be reflected back to asymptotic infinity. Depending on the actual value of $V_{\rm eff}|_{r=0}$, this leads to a significant gain of intensity for $b < b_{\rm crit}$. This effect can be used to rule out part of the Type Ic solution space based on shadow observations.

Finally, we exemplify the intensity profile associated with geometries of Type II in the right column of Fig.\ \ref{fig.intense}. In this case $V_{\rm eff}|_{r=0} = \infty$ and all light rays with $b > 0$ have a turning point and go out to asymptotic infinity. As a consequence, these geometries do not cast a shadow. Their profile resembles a star where the maximum intensity is reached in the center of the image. Based on this distinguished feature, Type II geometries are ruled out observationally.

\begin{figure}[t!]
	\centering
		\includegraphics[width = 0.6 \textwidth]{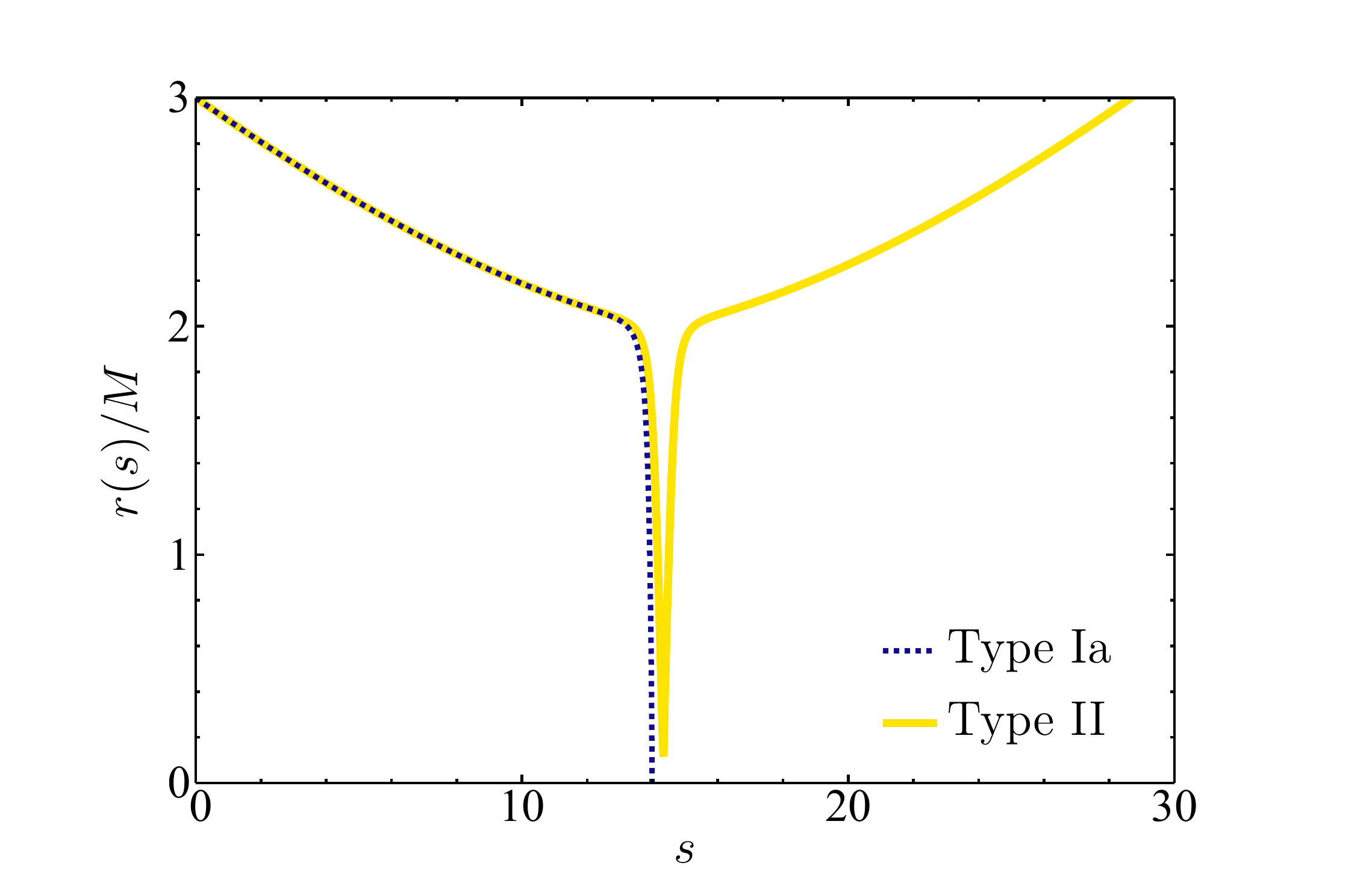} 
	\caption{\label{fig:exampleray} Radius $r(s)$ plotted as a function of the path parameter $s$, with an impact parameter of $b/M = 0.05$ for an object of mass $M=10$. For the geometry of Type Ia the ray terminates at the singularity while for the Type II geometry the light ray is reflected instead. This feature where the trajectory is extended, going back out to infinity again, is the underlying reason for why the brightness in the interior is enhanced for Type II and Type Ic solution, as can be seen in Fig.\ \ref{fig.intense}.}
\end{figure}

Heuristically, the enhanced brightness in the case of Type Ic and Type II solutions can be understood as follows. For all type of solutions, once a ray is inside the unstable photon region and headed towards the interior, the light ray will continue to head inwards and eventually approaches the singularity. For solutions of Type Ia or Ib (and similarly for Type III solutions) the light rays inevitably reach the singularity, and terminate there. For Type II solutions however, the light rays are reflected near the singularity due to the infinitely high potential barrier that is present at the center. The rays then extend outward to infinity again, following the mirrored trajectory but now in opposite direction, see Fig.\ \ref{fig:exampleray}. Due to this reflection, the intensity picked up by the ray is significantly enhanced for two reasons. First, the ray crosses the region where light is emitted twice, leading to an increase in the intensity. Secondly, light rays that first go towards the interior, are reflected, and then go out and reach the observer at infinity are first blueshifted, and then redshifted. The rather severe redshift effect obtained by climbing out of the interior is compensated by first falling inwards, negating its effect. For Type Ic solutions, only rays with an impact parameter above some critical value, determined by the height of the potential barrier, are reflected, thus still leaving a (smaller) shadow.

We point out that the unstable photon orbit imprints a peak at $b_{\rm crit}$ in the intensity profiles (Fig.\ \ref{fig.intense}). As the critical impact parameter agrees with the Schwarzschild value almost exactly, that peak position contains important information, the asymptotic mass of the object. For geometries of Type Ic and Type II, the peak provides an intrinsic scale in the image. 

\begin{figure}[t!]
	\centering
	\includegraphics[width = 0.31 \textwidth]{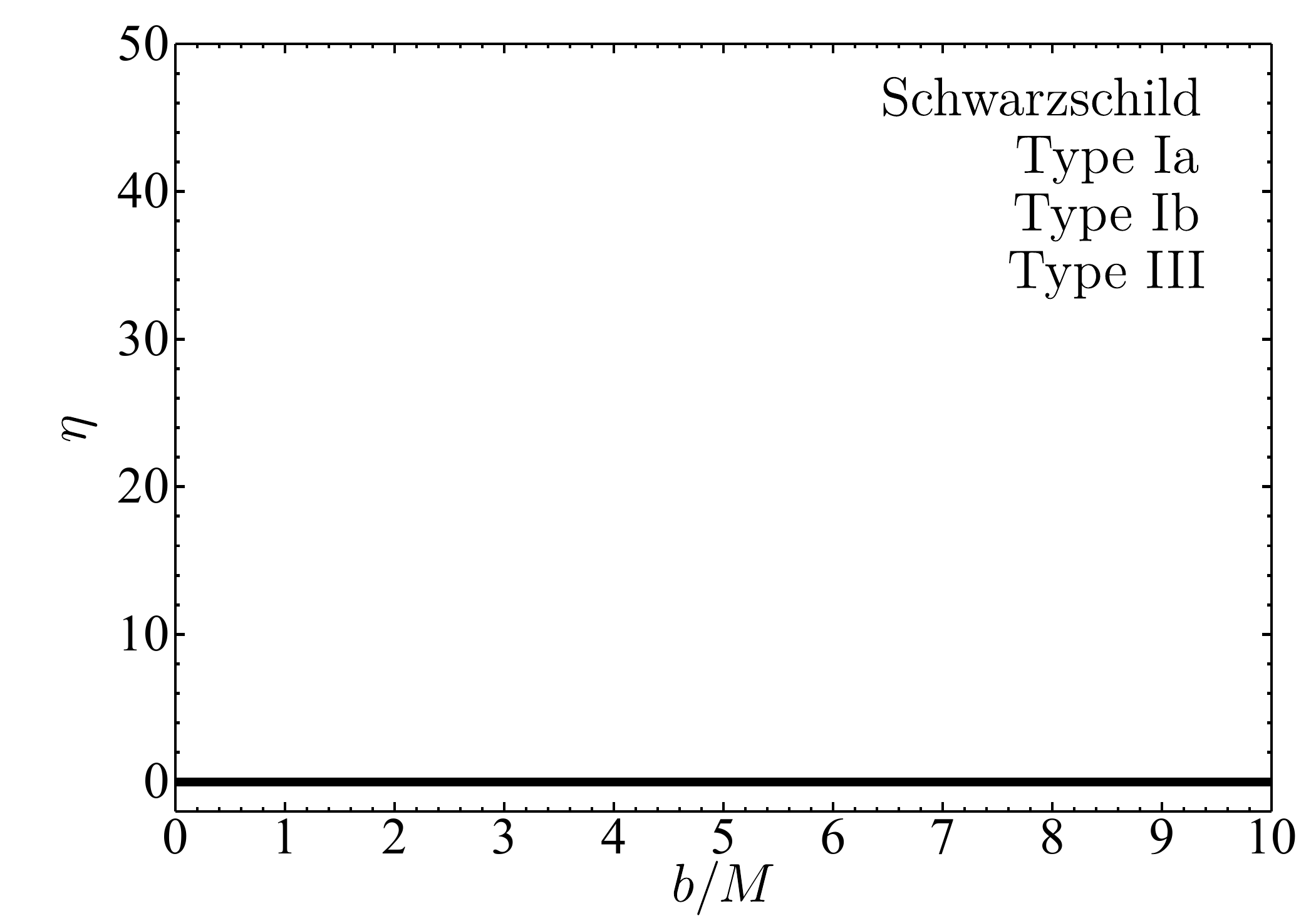} \,
	\includegraphics[width = 0.31 \textwidth]{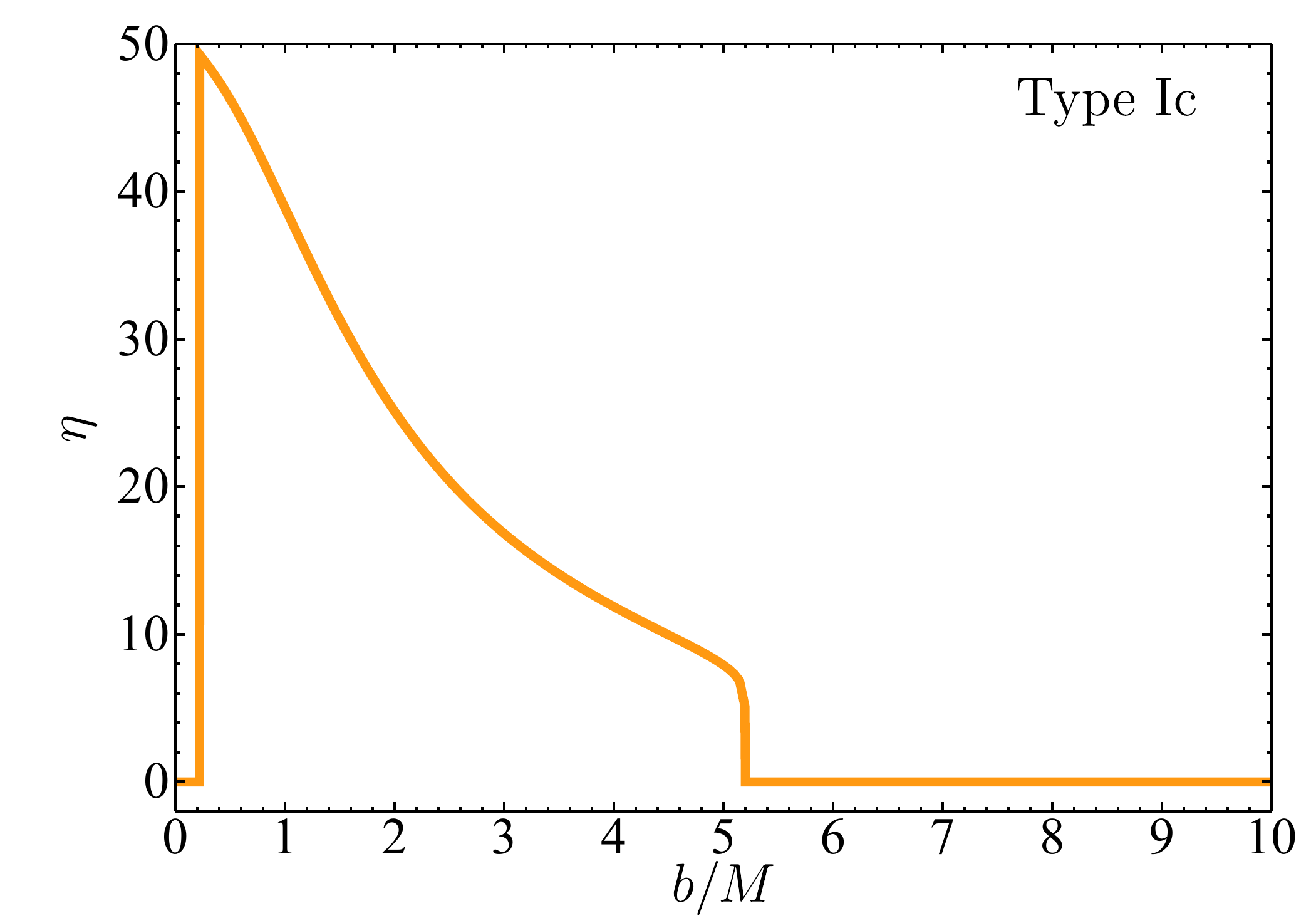} \, 
	\includegraphics[width = 0.31 \textwidth]{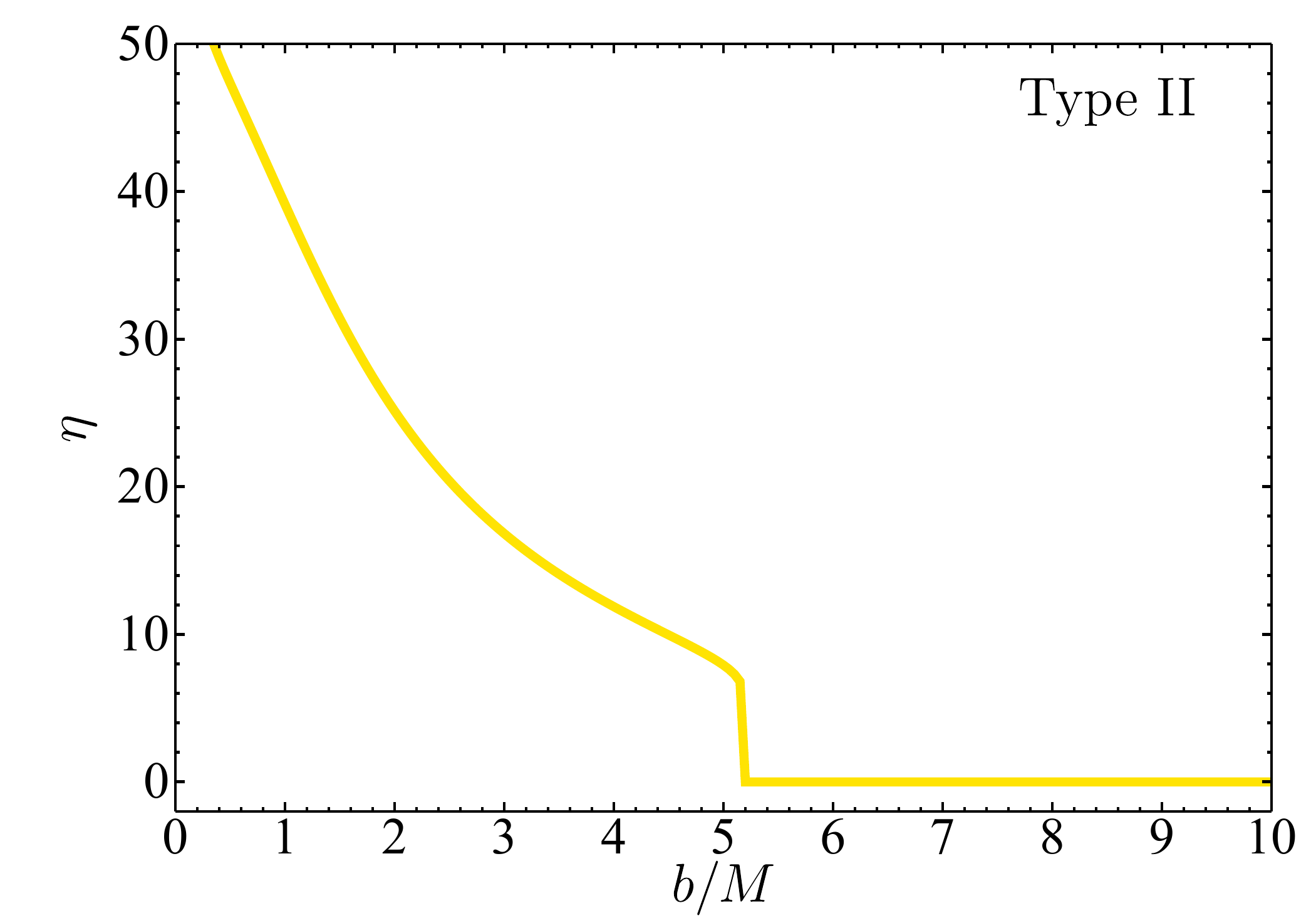} \; \; \;
	\caption{\label{fig.blip} Relative difference in intensity, eq.\ \eqref{eq:etadef} evaluated for geometries of Type Ia, Type Ib, and Type III (left), Type Ic (middle) and Type II (right) with mass $M=10$. The latter two cases show distinguished and strong deviations from the Schwarzschild case, allowing to distinguish between the geometries based on shadow images.}
\end{figure}
We complete our discussion by giving the relative strength of a signal distinguishing the Schwarzschild solution from alternatives, we define the relative difference in the intensity profiles 
\be\label{eq:etadef}
\eta \equiv \frac{I_{\rm model} - I_{\rm Schwarzschild}}{I_{\rm Schwarzschild}} \, .  
\ee 
Fig.\ \ref{fig.blip} then shows $\eta$ as a function of the impact parameter for all examples introduced in Table \ref{tab:initialconditions}. For geometries of Type Ia, Type Ib, and Type III, shown in the left panel, $\eta \ll 1$ and of the order of the numerical integration accuracy. Hence we do not give details for these cases. The associated regions in the quadratic gravity phase space, shaded in pink, red, and black in Fig.~\ref{fig:pspace}, are not excluded by black hole shadow observations consistent with the GR prediction. In contrast, the geometries of Type Ic and Type II exhibit distinguished and pronounced differences in the intensity profiles. For Type Ic, the size of the shadow depends on the specific parameter values, in particular on $\alpha$, $\beta$, $S_0^-$, and $S_2^-$, and hence the associated region in phase space can be constrained by more refined black hole shadow observations. For Type II, we find $\eta \approx \mathcal{O}(10)$ in the interior. The enhancement of the intensity toward the center of the image is a topological effect valid for all parameter combinations, in contrast to an intensity depression in the Schwarzschild case. This signature is sufficiently pronounced to completely rule out Type II as an alternative to the Schwarzschild black hole based on shadow observations~\cite{Akiyama:2019cqa}. Thus, at the level of Fig.~\ref{fig:pspace}, parts of the phase space shaded in orange can be restricted while yellow-shaded areas are not compatible with observations.

\section{Summary and Outlook}
\label{sec.4}
\subsection{Main Results}
This work focused on quadratic gravity, and its implications on images made by the EHT. In this course, we employed the following strategy. First we restricted to vacuum solutions that are static, spherically symmetric and asymptotically flat.  Then we looked at the asymptotic behavior of these solutions by solving the linearized equations of motion of quadratic gravity. This shows that,  at fixed Newton's coupling, there is a five-dimensional space of asymptotically flat geometries parameterized by the two additional coupling constants of the theory and three parameters spanning the solution space at fixed values of the couplings. The latter can be associated with the asymptotic mass $M$ of the solution and the strengths of the (exponentially suppressed) Yukawa-type interactions introduced by the additional massive degrees of freedom of the theory. The linearized solutions provide initial conditions for  integrating the \emph{full, non-linear} equations of motion numerically thereby obtaining global geometries. As their key feature, the global geometries track the Schwarzschild solutions almost perfectly for $r \gtrsim 2M$ while deviating substantially within the would-be event horizon.\footnote{A similar result, based on renormalization group improvements in the context of the gravitational asymptotic safety program has recently been reported in \cite{Borissova:2022jqj}.} The solutions are in agreement with the local scaling behaviors determined analytically by the Frobenius method. Our scan of the phase space, comprising around $10^4$ geometries, identified the following types of solutions. First, we recovered the Schwarzschild spacetime familiar from general relativity which is also a solution to quadratic gravity. Secondly, we identified naked singularities of Type I (coming in subclasses a,b,c) and Type II. Thirdly, we encountered wormhole solutions which are classified as Type III geometries. All solutions are compatible with the local scaling behaviors identified earlier by L\"u et.\ al.\ \cite{Lu:2015psa} and Holdom \cite{Holdom:2002xy}. Conversely, not all local scaling behaviors arising from the Frobenius analysis are realized by extending asymptotically flat solutions. Our work, for the first time, combines these results in a comprehensive picture on the phase space of static, spherically symmetric, and asymptotically flat vacuum solutions in quadratic gravity.

Our numerical investigation also supports the argument given in \cite{Lu:2015psa} that, for $\alpha \not =0, \beta \not = 0$ the Schwarzschild solution is the only geometry within the phase space exhibiting an event horizon. While it is known that the Stelle black hole provides an additional black hole geometry with a horizon \cite{Lu:2015cqa}, this does not appear in our phase space construction since our choice of $M$ is above the critical mass for which this branch of solutions exists. Thus, our analysis suggests the following Birkhoff-type uniqueness theorem for sufficiently massive black holes: \emph{The only static, spherically symmetric, and asymptotically flat solution of (generic) quadratic gravity compatible with the cosmic censorship hypothesis is the Schwarzschild solution.} 

Subsequently, we addressed the question whether images taken by the EHT can discriminate between the various types of geometries encountered in the phase space of quadratic gravity. In this context, the first observation made by our work is that all geometries constructed in Sect.\ \ref{sec.2} are essentially identical at coordinate distances $r \gtrsim 2M$ with significant deviations appearing close to and below $r = 2M$ only. In particular, the location of the unstable photon orbit setting the size of the shadow cast by the Schwarzschild geometry is identical up to exponentially small corrections. Thus the location of this orbit is unable to discriminate between the geometries.

Moreover, we equipped our geometries with the simple emission model proposed in Ref.\ \cite{Bambi:2013nla}.\footnote{In \cite{Bambi:2013nla}, this model was used to analyze intensity profiles related to phenomenologically motivated wormhole solutions. Later on, \cite{Shaikh:2018lcc} used the same model to construct characteristic intensity profiles associated with naked singularities of the JMN-1 and JMN-2 types. Depending on the model parameters, the resulting findings are in qualitative agreement with our results for Type I and Type II naked singularities.} It is shown that the resulting intensity profiles allow to easily distinguish the Schwarzschild geometry from geometries of Type Ic and Type II as these geometries exhibit a significant increase of intensity within the shadow region of the Schwarzschild black hole with identical asymptotic mass (c.f.\ Fig.\ \ref{fig.blip}). The  relative distribution of the effect is easily resolvable by the EHT. For realistic masses, these geometries essentially agree with the Schwarzschild solution outside the event horizon so that they cast identical shadows. Ultimately, this effect can be traced back to the fact that these geometries actually possess light-like geodesics with turning points at impact parameter $b < b_{\rm crit}$ where $b_{\rm crit}$ is the critical value separating the plunge orbits from rays going out to asymptotic infinity in the Schwarzschild geometry.
This feature is absent for the geometries of Type Ia, Type Ib, and Type III. As a consequence, the intensity profiles obtained in these cases are identical to the one of the Schwarzschild black hole for observational purposes.

These findings entail two important consequences. Firstly, the EHT has the capability of probing (and actually ruling out part of) the phase space of static, spherically symmetric, and asymptotically flat vacuum solutions in quadratic gravity. Secondly, this phase space contains horizonless naked singularities and wormhole geometries whose intensity profiles are virtually indistinguishable from the ones created by the Schwarzschild geometry. On this basis alone, it is difficult to conclude whether the object imaged by the telescope actually possesses an event horizon. 

\subsection{Outlook}
In this work, we established that observations by the Event Horizon Telescope can rule out parts of the phase space of quadratic gravity as alternatives to the black hole solutions found in general relativity. Clearly, the present work can be generalized in various directions. \\

\noindent
\emph{Geometries including angular momentum}. \\
Realistic black holes are expected to carry angular momentum. This applies  in particular to supermassive black holes in the center of galaxies. Although the inclusion of angular momentum has only a mild effect on the size and shape of the shadow cast by the black hole, slightly deforming the image asymmetrically, it would be highly interesting to generalize the geometries studied in the present work relaxing the assumption of spherical symmetry.  \\

\noindent
\emph{More realistic intensity profiles}. \\
A more detailed comparison between theoretical predictions and actual observations will require more elaborate synthetic images of the shadow. We expect that this is actually achievable by improving the accretion disk model and applying advanced ray tracing techniques developed by the EHT collaboration. The alternatives to the Schwarzschild geometry constructed in the present work may constitute theoretically well-motivated targets for such an investigation. In particular, it would be interesting to determine whether the improved emission model, potentially including outflows, actually allows to discriminate between the Schwarzschild solution and a naked singularity of Type Ib, putting observational constraints on this part of phase space as well. This will be investigated in a the companion paper \cite{inprep}.  \\

\noindent
\emph{Including quantum corrections}. \\
At this stage our work constructed the vacuum solutions of classical quadratic gravity. Since this theory is perturbatively renormalizable \cite{Stelle:1976gc}, it would be very interesting to study quantum corrections to the classical phase space by taking into account one-loop corrections to the action \eqref{eq:qg}. Similarly, it would be interesting to understand how the two-loop counterterm found by Goroff and Sagnotti \cite{Goroff:1985th} modifies the vacuum solutions of general relativity. Since some of the intensity profiles constructed in our work actually probe spacetime in a regime where (quantum) gravity effects are expected to be strong it is actually conceivable that there could be signatures of such terms contained in the observations. \\

\noindent
\emph{Improving the image resolution}. \\
 In comparison to the Schwarzschild solution, the shadow size of Type Ic geometries is decreased. Additionally, there is a ring-like brightness excess region appearing at the horizon position in the Schwarzschild case. This effect goes beyond a pure change of the shadow size, it also modifies the intensity profile. Depending on the parameters specifying the solution, both features can be more or less pronounced and might be limited to a small angular scale. Thus they may be beyond the current effective resolution of the EHT which amounts to $\sim$25\,$\mu$as or roughly half of the ring diameter in the case of M87$^\ast$ \cite{EventHorizonTelescope:2019PaperI}. However, it might be feasible in the future when extending the size of the Earth-sized telescope array into space by using satellites which improves the EHT resolution by one order of magnitude or more \cite{Roelofs:2019nmh}. In addition, upcoming observations related to Sgr A$^\ast$ may provide further observational constraints. \\

\noindent
\emph{Additional observables}. \\
Our present discussion focused on discriminating geometries based on observation channels related to the EHT. In a similar spirit, \cite{Salvio:2022mld} aimed to constrain quadratic gravity based on early universe cosmology. Clearly, it would be interesting to complement these works by other observational channels related to, e.g., the emission of gravitational waves. This may provide constraints which are  complementary to the ones based on shadow observations, allowing us to further constrain deviations from general relativity motivated by fundamental quantum gravity considerations.

\section*{Acknowledgments}
We thank M.\ Ba{\~n}ados, A.\ Bonanno, T.\ Bronzwaer, and H.\ Olivares for interesting discussions and A.\ Khosravi and M.\ Galis for participating in the early stages of this investigation. Furthermore, we thank C.\ Bambi for correspondence on earlier work \cite{Bambi:2013nla}. This project was enabled by the Radboud Excellence fellowship awarded to M.F.W.\ by Radboud University in Nijmegen, Netherlands. The work of F.S.\ is supported by the NWA-grant ``The Dutch Black Hole Consortium''. \\[2ex]

\noindent
While this work was under internal review, two preprints with content related to the solution classification appeared~\cite{Silveravalle:2022lid,Bonanno:2022ibv}.\\


\end{document}